\documentclass[]{interact}
\usepackage{epstopdf}
\usepackage{natbib}
\bibpunct[, ]{(}{)}{;}{a}{}{,}

\usepackage{upgreek}
\usepackage{url}
\usepackage{hyperref}
\usepackage{breakurl}
\usepackage{color}
\usepackage{multicol}
\def\pvint{\mskip5mu \raise0.5ex \hbox{\vrule height0.5pt depth0.3pt
 width0.2cm} \hskip-9.3pt\int}
\def\textpvint{\mskip1mu \raise0.5ex \hbox{\vrule height0.6pt depth0.0pt
 width0.16cm} \hskip-7.1pt\int}
\def\rset{\hbox{{I\kern -0.2em R}}}
\def\cset{\hbox{{C\kern -0.55em\raise 0.5ex\hbox{{\tiny $|$}}}}}
\def\iset{\hbox{{I\kern -0.2em I}}}

\newcommand{\ue}{\mathrm{e}}
\newcommand{\ui}{\mathrm{i}\,}

\renewcommand{\pi}{\uppi}

\begin{document}

\title{Exact solutions for radiative transfer with partial frequency
  redistribution}
\author{
\name{H.~Frisch\textsuperscript{a}\thanks{CONTACT H.~Frisch.
      Email: frisch@oca.eu To be published in the Journal of Computational
      and Theoretical Transport, DOI 10.1080/23324309.2024.2393598}}
 \affil{\textsuperscript{a}Universit\'e C\^ote d'Azur, Observatoire de la
  C\^ote d'Azur, CNRS, Laboratoire Lagrange, Bd de
  l'Observatoire, CS34229, 06304 Nice Cedex 4, France}
}
\maketitle

\begin{abstract}

The construction of exact solutions for radiative transfer in a
plane-parallel medium has been addressed by Hemsch and Ferziger in
1972 for a partial frequency redistribution model of the formation of
spectral lines consisting in a linear combination of frequency
coherent and fully incoherent scattering. The method of solution is
based on an eigenfunction expansion of the radiation field, leading to
two singular integral equations with Cauchy or Cauchy-type kernels, that have
to be solved one after the other. 
We reconsider this problem, using as
starting point the integral formulation of the radiative transfer
equation, where the terms involving the coupling between the two
scattering mechanisms are clearly displayed, as well as the primary
source of photons.
 
With an inverse
Laplace transform, we recover the singular integral equations
previously established and, with Hilbert transforms as in the previous
work, recast them as boundary value problems in the complex plane.
Their solutions are presented in detail for an infinite and a
semi-infinite medium. The coupling terms are carefully analyzed
and consistency with either the coherent or the incoherent limit is
systematically checked.

We recover the important result of the
previous work that an exact solution exists for an infinite medium,
whereas for a semi-infinite medium, which requires the introduction of
half-space auxiliary functions, the solution is given by a Fredholm
integral equation to be solved numerically. The solutions of the
singular integral equations are used to construct explicit expressions
providing the radiation field for an arbitrary primary source and for
the Green's function. An explicit expression is given for the radiation
field emerging from a semi-infinite medium. 
\end{abstract}


\section{Introduction}
\label{sec-intro}
It has been known since the work of the mathematicians N.~Wiener and
E.~Hopf in 1931 \citep{wiener31} that monochromatic scattering in a 
semi-infinite medium has an exact solution. Monochromatic means that
scattering remains frequency coherent, photons suffering at each
scattering only changes in direction. The so-called Wiener--Hopf method
of solution is based on the possibility to express the radiative
transfer equation as a convolution integral equation on a half-line
for a function depending only on the space variable. 

For incoherent scattering, that is when the frequencies of photons before
and after a scattering are totally uncorrelated, it was shown,  thanks to the
work done in the late forties and early fifties by the group of
V.V.~Sobolev in Leningrad
\citep[for references see e.g.][]{sobolev63,ivanov73}, that the radiative
transfer problem can also be recast as a convolution integral equation
for a function which depends only on the space variable.  
The Wiener--Hopf method remains applicable, as rigorously proved by
\citet{krein62} \citep[see also][]{levinson49}. 
In Astronomy, this type of scattering, which applies to most of the
spectral lines observed in the spectra of stellar atmospheres, is
usually referred to as {\em complete frequency redistribution}, a 
terminology that is followed here.  In contrast, monochromatic
scattering applies to the formation of continuous spectra emitted by
stellar atmospheres.

For spectral lines, the assumption of complete frequency
redistribution breaks down, when 
the frequencies of the absorbed and scattered photons are correlated.
\emph{ Partial frequency redistribution}, as it is called, is typical
of strong spectral lines, known as resonance lines (see
Appendix~\ref{sec-kneer} for details). In principle it is not possible to 
find exact solutions when there is partial frequency redistribution
because the radiative transfer equation cannot be recast into an
equation for a function depending only on the space variable. However,
for a very 
simple partial frequency redistribution model, consisting in a linear
combination of monochromatic scattering and complete frequency
redistribution, it was shown by M.~Hemsch in his PhD thesis with a
method based on the singular eigenfunction expansion method introduced
 in \citet{case60},
that it is possible to construct an exact solution for a infinite
medium and to reduce the problem to a Fredholm integral equation for a
semi-infinite medium \citep{{hemsch71},hemsch72}. This simple partial
frequency redistribution model was introduced for resonance lines (see
the discussion in Appendix~\ref{sec-kneer}), for spectral lines formed
between two excited levels 
\citep[p.~252]{jefferies68, frisch80, hubeny15, frisch22} and to study
the resonance scattering  of gamma-ray quanta \citep{yengi91}.
The same type of model is found in \citet{corngold67} to analyze the
decay of a neutron pulse inside a moderator.  Several examples of
applications to neutron and photons scattering problems of the
expansion method introduced in \citet{case60} 
may be found in classical textbooks and reviews
\citep[e.g][]{casezwei67,williams71,cormick73,duder79} 
A few of them are also mentioned at the end of Section~\ref{sec-transfereq}. 

Because of the interplay between monochromatic scattering and complete
frequency redistribution, the solution is not a simple linear
combination of a monochromatic scattering and a complete frequency
redistribution solution. As shown in \citet{hemsch71} and
\citet{hemsch72}, a solution can be constructed by proceeding in two
successive steps.  The first one consists in solving a monochromatic
scattering problem. The integration over frequency of its solution
provides a generalized complete frequency redistribution problem,
involving only frequency integrated functions. This two-step procedure
is also applied in \citet{yengi91}, where the method of solution is
otherwise radically different from the method employed in
\citet{hemsch71} and the results not as explicit.

In this article, we also consider a linear combination of
monochromatic scattering and complete frequency redistribution, 
reach the same conclusions as in \citet{hemsch71} and
\citet{hemsch72}, but with an approach that has differences and similarities 
with the previous work. In  the latter a
singular eigenfunction expansion method is used to transform the
integro-differential equation for the radiation field 
into a Cauchy-type singular integral equation, which has the
particularity of having two singular integrals, one coming from the
coherent scattering contribution and the other one from the incoherent
one.  Assuming that the incoherent contribution is
a known term, the singular integral reduces to a singular
integral equation for monochromatic scattering. An integration over
frequency of its solution provides a new singular integral equation
for a frequency independent function, the properties of which
are akin to those of a singular integral equation for complete
frequency redistribution. The solutions of both equations are obtained
by transforming them  into a boundary value problem in the  complex
plane. This method of solution, proposed by \citet{carleman22}, relies
on the analyticity properties of Hilbert transforms 
\citep[e.g.,][]{musk53,gakhov66,pogo66,ablowitz97}.

The hard core of our approach is also the solution of a
singular integral equation containing two singular integrals, coming from
the coherent and incoherent parts respectively, and its solution is
constructed in two steps, as described above. The route
to the singular integral is however different. The radiative transfer
equation is first transformed into a convolution integral equation for
the direction averaged radiation field.  The latter is then transformed
into a singular integral equation by an inverse Laplace
transform, more or less equivalent to the singular eigenfunction
expansion. Since the convolution equation takes into account the 
inhomogeneous term in the radiative transfer equation and the boundary
conditions on the radiation field, the solution of the singular
integral equation directly provides the full solution of the
problem. In contrast, the 
singular integral equation derived with the singular eigenfunction
expansion provides only a general solution of the homogeneous
radiative transfer equation, to which must be added a particular
solution of the inhomogeneous equation. The full solution has then to be
constrained by the boundary conditions.

The organization of this work is as follows. The radiative transfer
and its two associated integral equations are constructed in
Section~\ref{sec-transfer}. Section~\ref{sec-dispersionfreq} is
devoted to the {\em monochromatic dispersion function}, which plays a
critical role in the solution of monochromatic scattering problems.
In Section~\ref{sec-singular} we consider an infinite medium and give
an explicit solution of the infinite medium Green's function. In
Section~\ref{sec-half}, we consider a semi-infinite medium and show
how to construct the half-space auxiliary functions, corresponding
respectively to the monochromatic and the generalized complete
frequency redistribution problem.  We also show how to calculate the
emergent radiation field. Some remarks and technical material are
presented in Appendices. In Appendix~\ref{sec-kneer} we discuss the
modelling of resonance lines by a linear combination of monochromatic
scattering and complete frequency redistribution. In
Appendix~\ref{sec-whtocauchy} we explain how a convolution integral
equation can be transformed into a singular integral
equation. Appendix~\ref{sec-omega} is devoted to the properties of the
dispersion function associated to the generalized complete frequency
redistribution problem. Finally, Appendix~\ref{sec-jump} 
presents some details regarding the construction of the half-space
auxiliary functions.

\section{The radiative transfer equation and associated integral equations}
\label{sec-transfer}
We first present the radiative transfer equation for a
combination of coherent and incoherent scattering. We then show how to
transform it into a convolution integral equation and then how the latter
can then be transformed into a singular integral equation.

\subsection{The radiative transfer equation}
\label{sec-transfereq}
We consider here an isolated spectral line. Following a notation
employed in \citet{chandra60} and a series of 
articles by D.~Hummer and collaborators \citep[see e.g.][]{avrett65},
we write the one-dimensional radiative transfer equation as
\begin{equation}
\mu\frac{\partial I}{\partial
  \tau}(\tau,x,\mu)=\varphi(x)[I(\tau,x,\mu) -S(\tau,x)],  
\label{eq-transfer}
\end{equation}
where $I(\tau,x,\mu)$ is the specific intensity of the radiation field
and $S(\tau,x)$ the so-called source function. For the problem
considered here, it is defined in Eq.~(\ref{def-source}). Here $\tau$
is the frequency-averaged line optical depth. Its domain of definition
is $\tau\in]-\infty,+\infty[$ for an infinite medium and
 $\tau\in[0,+\infty[$\footnote{The notation $[\ ]$ stands for a
 closed interval, $[\ [$ or $]\ ]$ for semi-open intervals
 and $]\ [$ for an open interval such as
 $]-\infty,+\infty[$.} for a semi-infinite one. The
 variable $\mu\in[-1,+1]$ is the direction cosine of the
 photon-propagation vector, as measured from the negative
 $\tau$ axis. With this choice, usual in Astronomy, $\mu$ is
 positive for rays emerging from the surface of a
 semi-infinite medium, negative for incoming rays, and the
 absorption term appears with a positive sign in the right-hand side.
        The variable $x\in]-\infty,+\infty[$ measures frequencies
        from the line center in a convenient
        unit, usually the Doppler width of the line, assumed to be
        independent of $\tau$.  The line absorption profile
        $\varphi(x)$ is normalized to unity in $-\infty\le x\le
        +\infty$.  For spectral lines, $\varphi(x)$ is usually assumed
        to be a Gaussian or a Voigt function (convolution of a
        Gaussian with a Lorentzian).  Here we assume that the source
function $S(\tau,x)$, which describes the emission term, has the form
\begin{equation} 
  S(\tau,x)= (1-\epsilon)[bJ(\tau,x) + (1-b)\bar J(\tau)] + Q^\ast(\tau),
\label{def-source}
\end{equation}
where $Q^\ast(\tau)$ is a given primary source of photons,
\begin{equation}  
J(\tau,x)=\frac{1}{2}\int_{-1}^{+1}
I(\tau,x,\mu)\,d\mu,\quad \mbox{and}\quad\bar J(\tau)=
\int_{-\infty}^{+\infty}J(\tau,x)\varphi(x)\,dx.  
\label{def-means}
\end{equation}
The parameter $\epsilon\in[0,1]$ is the probability that upon
scattering a photon is lost from the line. We shall also make use of
$\omega=1-\epsilon$ the albedo for single scattering. 
The coefficient $b\in[0,1]$ is the branching ratio between
monochromatic scattering and complete frequency redistribution, the limit
$b=0$ corresponding to complete frequency redistribution and the limit $b=1$
to monochromatic scattering.  In this limit, the scattering
of photons occurring without changes in frequencies, the
dependence of $I(\tau,x,\mu)$ on the frequency $x$ can be ignored and
$\varphi(x)$ set to one. In both limits the source function depends
only on $\tau$.
To make a contact with particle transport, we remark here that
  $I(\tau,\mu,x)$ 
corresponds to a one-particle distribution function $f(\bm r,\bm v)$,
and that the radiative transfer equation is a time-independent linear Boltzmann
equation. 

In \citet{hemsch72}, the branching ratio is assumed to be a function
$b(x)$ of the frequency. Here we keep it independent of $x$, for a
fundamental reason, namely that a branching ratio depending on $x$
does not preserve conservation of energy nor symmetry in $x$ and $x'$
(see Appendix~\ref{sec-kneer} for a more detailed argumentation).
The assumption that $b$ is a constant does not influence the main
conclusions of the work of \citet{hemsch72}.

In \citet{hemsch72},  it is  assumed that the
solution of the radiative transfer equation, restricted to its homogeneous
part, may be written as
\begin{equation}  
I(\tau,x,\mu)=\int_k\Phi(k,x,\mu)\,\ue ^{-\tau/k}\,dk.   
\label{eq-hemschferziger72}
\end{equation}
This expansion, once inserted into the homogeneous part
of Eq.~(\ref{eq-transfer}) provides a relation between $\Phi(k,x,\mu)$ and
the two  averages $\bar \Phi(k,x)$ and $\bar{\bar \Phi}(k)$, defined as in
Eq.~(\ref{def-means}).  The condition that this expansion, say at
$\tau=0$, can represent an arbitrary function $\psi(x,\mu)$, provides
a integral equation with two Cauchy-type singular integrals, one for
$\bar \Phi(k,x)$ and one for $\bar{\bar\Phi}(k)$. Its solution is
carried out in two steps. The first one
provides $\bar \Phi(k,x)$ in terms of $\bar{\bar\Phi}(k)$. An
integration over frequency of $\bar 
\Phi(k,x)$ provides a new equation, now of the complete frequency
redistribution type, which can be solved for $\bar{\bar
  \Phi}(k)$.  Once $\Phi(k,x,\mu)$ has been
determined, it can used to construct a solution of the inhomogeneous
radiative transfer equation, which takes into
account the boundary conditions.

The proof given in \citet{hemsch71} and \citet{hemsch72} that exact
solutions can be constructed for a partial frequency redistribution
problem is definitely a great success of the singular eigenfunction
expansion method. Another example is the proof by \citet{siewert67}
that conservative Rayleigh scattering in a semi-infinite medium has an
exact solution. The first proof of this result was given by
\citet{chandra46} \citep[see also][]{chandra60} with a method in which
the most important point is somewhat hidden. With a singular
eigenfunction expansion, \citet{siewert67} show how the vectorial
transfer equation for the linearly polarized radiation field can be
transformed into a matrix boundary value problem in the complex
plane \citep[see also][]{siewert72}. The latter has an exact solution
because the associated matrix 
dispersion function can be diagonalized.  This property breaks down
when the scattering looses the monochromatic and/or its conservative
characteristics. More generally, the singular eigenfunction expansion
method brings a rigorous and deep mathematical understanding to transport 
problems involving the scattering of neutrons or photons and as such
offers an excellent starting point for the construction of accurate
numerical solutions of neutron and photons transfer
equations\citep[e.g.][]{siewert79,barichello98,barichello99a,barichello99b}.

Now, as explained in Section~\ref{sec-intro} we shall construct a
convolution integral equation with an assumption somewhat similar
to Eq.~(\ref{eq-hemschferziger72}) and then transform it into a singular
integral equation with two singular integrals. This method is
applied in Section~\ref{sec-singular} for an infinite medium and in
Section~\ref{sec-half} for a semi-infinite one.

\subsection{Convolution integral equations}
\label{sec-convolution}

We show here how a radiative transfer equation can be transformed into
a convolution integral equation for the source function and mention
some of the advantages of this formulation. For monochromatic
scattering, the systematic introduction of of convolution integral
equations for handling radiative transfer equations dates back from
\citet{milne21} and for complete frequency redistribution from the
late forties and early fifties, thanks to the work done by the group
of V.V.~Sobolev in Leningrad \citep[for references see
  e.g.][]{sobolev63,ivanov73}. The integral equation for the source
function is obtained by inserting a formal solution of the radiative
transfer equation into the definition of the source function.

Let us consider for example a semi-infinite medium with no incident
radiation on the surface and complete frequency redistribution. The
source function is given by 
$S(\tau)=(1-\epsilon)\bar J(\tau) + Q^\ast(\tau)$, where $\bar
J(\tau)$ is the direction and frequency average of the radiation field
(see Eq.~(\ref{def-means})). 
For $\mu>0$, the formal solution of the radiative transfer may be written as
\begin{equation}  
  I(\tau,x,\mu)=\int_\tau^{\infty} S(\tau')\,
  \ue ^{-[\tau'-\tau]\varphi(x)/\mu}\frac{\varphi(x)}{\mu}\,d\tau'.
\label{eq-soltransplus} 
\end{equation}
There is for $\mu<0$ a similar expression, in which the integration runs 
from $0$ to $\tau$. Integrating Eq.~(\ref{eq-soltransplus}) over $\mu$ and
over $x$ (weighted by $\varphi(x)$), we readily obtain the integral equations  
\begin{equation}  
  S(\tau)=(1-\epsilon)\int_0^{\infty}K(|\tau-\tau'|)
S(\tau')\,d\tau' + Q^\ast(\tau),
\label{eq-convolrc}
\end{equation}
and
\begin{equation}  
\bar J(\tau)=(1-\epsilon)\int_0^{\infty}K(|\tau-\tau'|)
\bar J(\tau')\,d\tau' + \int_0^{\infty}K(|\tau-\tau'|)Q^\ast(\tau')\,d\tau'.  
\label{eq-convoljtau}
\end{equation}
The kernel $K(\tau)$ is given by
\begin{equation}  
K(\tau)= \frac{1}{2}\int_{-\infty}^{\infty}
\int_0^1\varphi^2(x)\ue^{-|\tau|\varphi(x)/\mu}
\frac{d\mu}{\mu}\,dx.
\label{eq-defkernelrc}
\end{equation}
    
For monochromatic scattering, we can ignore the dependence on $x$ and
set $\varphi(x)=1$. The source function may then be written as
$S(\tau)=(1-\epsilon)J(\tau) + Q^\ast(\tau)$, where $J(\tau)$ is the
radiation field averaged over $\mu$. The functions $S(\tau)$ and $J(\tau)$
also satisfy Eqs.~(\ref{eq-convolrc}) and (\ref{eq-convoljtau}). The
kernel is given by 
\begin{equation}  
  K(\tau)=\frac{1}{2}\int_1^\infty
  \ue^{-|\tau|/\mu}\,\frac{d\mu}{\mu}. 
\label{kernel-mono}
\end{equation}
For monochromatic scattering, the kernel is one-half of the first
integro-exponential function. It is easy to observe that it can be
written as a Laplace transform, namely as 
\begin{equation}  
  K(\tau)=\int_0^\infty k(\nu)\,\ue ^{-\nu|\tau|}\,d\nu,
\label{eq-kernellaplace}
\end{equation}
where $k_{\rm M}(\nu)$ ($M$ for monochromatic) can be defined as
\begin{equation}
  k_{\rm M}(\nu)\equiv {1}/{2\nu},\,\,\nu>1, \quad k_{\rm M}(\nu)\equiv
  0,\,\,0<\nu<1. 
  \label{def-knumono}
\end{equation}
As stressed by V.V. Sobolev \citep[e.g.][]{sobolev63,ivanov73}, for
complete frequency redistribution, when $\varphi(x)/\mu$ is chosen as
integration variable, $K(\tau)$ can also be written as in
Eq.~(\ref{eq-kernellaplace}).  The definition of $k_{\rm R}(\nu)$ (R
for complete frequency redistribution) is given in
Eq.~(\ref{eq-knurc}).

Because they are constructed with a solution,
    albeit formal, of the radiative transfer equation, integral
    equations can take into account any boundary conditions on the
    radiation field. For example, an incident field $I_0(0,x,\mu)$,
    $\mu<0$, will appear as an additional primary source describing
    the directly transmitted incident radiation, provided the diffuse
    field is separated from the directly transmitted one. Here we
    always assumed for simplicity that the incident field on the
    boundary is zero and that there is an internal source.
    
The integral equations in Eqs.~(\ref{eq-convolrc})
      and (\ref{eq-convoljtau}) 
are the starting points for two of the most famous methods providing
exact solutions for radiative transfer in a semi-infinite medium. For
monochromatic scattering, when
$\epsilon=0$ and $Q^\ast(\tau)=0$, Eq.~(\ref{eq-convolrc}) becomes the
Milne integral equation, for which \citet{wiener31} constructed an exact
solution. Actually, the Milne integral was first proposed by  
\citet{schwarz14} to solve the problem stated 
in \citet{schwarz06} of the run of temperature in a stellar atmosphere
in radiative equilibrium. The Wiener--Hopf method of solution is based
on the transformation of the semi-infinite convolution integral
equation into a convolution equation on a full line. The use of the
analyticity properties of complex Fourier transforms leads to a
factorization problem in the complex plane. The Wiener–-Hopf method,
introduced for monochromatic transfer, can also be applied to complete
frequency redistribution, as was rigorously shown by
\citet{krein62}. This generalization makes clear the strict analogy
between a Wiener--Hopf factorization and a homogeneous
Riemann--Hilbert problem. Because it requires only a factorization in
the complex plane, the Wiener--Hopf method has been applied to
convolution integral equations with a large variety of kernels
\citep[see for
  example][]{noble58,krein62,pincus81,ablowitz97,lawrie07}. For
mono-kinetic neutron transport, it was very popular in the early
forties, but the publications came out only after 1945
\citep[e.g.][]{mark47,placzek47,seidel47}.

The fundamental work of V.~A.~Ambarzumian for monochromatic scattering
in the early forties is also based on the integral equation for the
source function \citep{ambar42,ambar43,kour63}.  When $Q^\ast(\tau)$
is replaced by the Dirac distribution $(1-\epsilon)\delta(\tau)$, the
inhomogeneous term in Eq.~(\ref{eq-convoljtau}) becomes
$(1-\epsilon)K(\tau)$ and the solution of Eq.~(\ref{eq-convoljtau})
provides the so-called \emph{resolvent function}, which is the regular
part of the surface Green’s function. Making use of the property that
the kernel and the inhomogeneous term are Laplace transforms,
Ambarzumian could establish that the resolvent function can be
expressed in terms a half-space auxiliary function, widely known now
as the H-function, a name attributed to it by S.~Chandrasekhar in the
late forties, and show that the H-function satisfies a nonlinear
integral equation.  This approach, often referred to as the resolvent
method, has been generalized to complete frequency redistribution
\citep[e.g.][]{sobolev63,ivanov73,ivanov94}. It does not invoke
analyticity properties of functions of a complex variable.
The resolvent method is applied to the partial 
frequency redistribution problem investigated by \citet{yengi91}.

For partial frequency redistribution, it is also possible to write an
integral equation for the function $S(\tau,x)$, but in general, it can
be solved numerically only. In the special case of a linear
combination of monochromatic scattering and complete frequency
redistribution, it is possible, as we now show, to construct for
${J}(\tau,x)$ an integral equation somewhat similar to
Eq.~(\ref{eq-convoljtau}), which will be the starting point for the
method of solution presented in this article.

The radiative transfer equation in Eq.~(\ref{eq-transfer})  shows that
the dependence on the frequency $x$ is actually through the absorption
profile $\varphi(x)$. The frequency variable is from now on
\begin{equation}  
  \xi=\varphi(x).  
\label{eq-variables}
\end{equation}
The variable $\xi$ is positive
and its range of variation is $[0,\varphi_0]$, with
$\varphi_0=\varphi(0)$. For the radiation field and the source
function, we use the notation $I(\tau,\xi,\mu)$ and $S(\tau,\xi)$ and
for the direction and frequency averaged intensity we introduce
${\cal J}(\tau,\xi)= J(\tau,x)$ and $\bar {\cal J}(\tau)$. The latter
is defined by  
\begin{equation}  
 \bar {\cal J}(\tau)=\bar
 J(\tau)=\int_{-\infty}^{+\infty}\varphi(x)J(\tau,x)\,dx=  
\int_0^{\varphi_0}W(\xi){\cal J}(\tau,\xi)\,d\xi,
\label{def-intjbar}
\end{equation}
where
\begin{equation}  
   W(\xi)=2\xi|\frac{dx}{d\xi}|.
\label{def-wxi}
\end{equation}
The normalization of $\varphi(x)$ to one implies
\begin{equation}  
 \int_0^{\varphi_0}W(\xi)\,d\xi=1. 
\label{eq-normewxi}
\end{equation}
The radiative transfer equation becomes
\begin{equation}
\mu\frac{\partial I}{\partial
  \tau}(\tau,\xi,\mu)=\xi[I(\tau,\xi,\mu) - S(\tau,\xi)].  
\label{eq-transferxi}
\end{equation}
The source function $S(\tau,\xi)$ is defined in
Eq.~(\ref{def-source}), with ${\cal J}(\tau,\xi)$ and  $\bar{\cal J}(\tau)$
instead of $J(\tau,x)$ and $\bar J(\tau)$. Integrating over $\mu$ the
formal solution given in 
Eq.~(\ref{eq-soltransplus}), in which $S(\tau')$ is replaced by
$S(\tau',\xi)$, we obtain for ${\cal J}(\tau,\xi)$ the integral equation 
\begin{eqnarray}  
 & & {\cal J}(\tau,\xi)=(1-\epsilon)\,b\,\xi\int_{-\infty}^{+\infty} K_{\rm
  M}(|\tau'-\tau|\xi){\cal J}(\tau',\xi))\,\,d\tau'\nonumber\\ 
& & + \ \xi \int_{-\infty}^{+\infty}K_{\rm M}(|\tau'-\tau|\xi)\,[Q^\ast(\tau')  +
  (1-\epsilon)(1-b)\bar{\cal J}(\tau')]\,\,d\tau'.
\label{eq-avintbis2}
\end{eqnarray}
It is written here for an infinite medium, but has exactly the same form
for a semi-infinite medium with no incident radiation. 

 The subscript M indicates that the kernel is the monochromatic one
 defined in Eq.~(\ref{kernel-mono}). For $b=1$, the term depending on
 $\bar{\cal J}(\tau)$, which describes the coupling between the two
 scattering mechanisms, disappears and when in addition $\xi$ is set to
 one, we recover the integral equation for $\bar J(\tau)$ written in
 Eq.~(\ref{eq-convoljtau}).  The main interest of
 Eq.~(\ref{eq-avintbis2}) is that $\bar{\cal J}(\tau)$ can be treated
 exactly as a primary source similar to $Q^\ast(\tau)$.  This property
 is fully exploited for the solution of the singular integral equation
 constructed in Section~\ref{sec-transformation}. We note that an
 integration of Eq.~(\ref{eq-avintbis2}) over $\xi$, weighted by
 $W(\xi)$, does not provide an integral equation for $\bar {\cal
   J}(\tau)$, because of the structure of the first integral in the
 right-hand side.

\subsection{Construction of a singular integral equation}
\label{sec-transformation}

There are different methods to transform a radiative transfer equation or a
convolution integral equation into a singular integral equation. Some
of them are mentioned  in the Appendix~\ref{sec-whtocauchy}. Here we
apply to Eq.~(\ref{eq-avintbis2}) an inverse Laplace transform method,
which amounts to assume that ${\cal J}(\tau,\xi)$ can be written as a
Laplace transform. For an infinite medium, we write
\begin{equation}  
  {\cal J}(\tau,\xi)=\left\{\begin{array}{cc}
\int_0^\infty j(\nu,\xi)\,\ue ^{-\nu\tau}\,d\nu & \quad \tau\ge 0,\\
 & \\
-\int_{-\infty}^0j(\nu,\xi)\,\ue ^{-\nu\tau}\,d\nu & \quad
\tau\le 0.\end{array}\right.  
\label{eq-inverse1}
\end{equation}
For a semi-infinite medium, it suffices to keep the positive values of $\tau$.
In Eq.~(\ref{eq-inverse1}) $\nu$ is real and $j(\nu,\xi)$, henceforth referred
to as the inverse Laplace transform of ${\cal J}(\tau,\xi)$, is a
regular function or a distribution. We also introduce the inverse Laplace
transforms of $\bar {\cal J}(\tau)$ and $Q^\ast(\tau)$, denoted
${\bar\jmath}(\nu)$ and $q^\ast(\nu)$ and their linear combination
\begin{equation}
  p^\ast(\nu)\equiv \omega(1-b)\bar\jmath(\nu) + q^\ast(\nu),
\label{eq-defpstar1}
\end{equation}
where $\omega$, the single scattering albedo, is defined by
\begin{equation}  
  \omega\equiv 1-\epsilon,
\label{eq-defvarpi}
\end{equation}
and ${\bar {\jmath}}(\nu)$ is related to $j(\nu,\xi)$ by
\begin{equation}  
  {\bar {\jmath}}(\nu)=\int_0^{\varphi_0}W(\xi)\,j(\nu,\xi)\,d\xi.
\label{def-barjnu}
\end{equation}

\begin{figure}
\begin{center}
\includegraphics[scale=0.4]{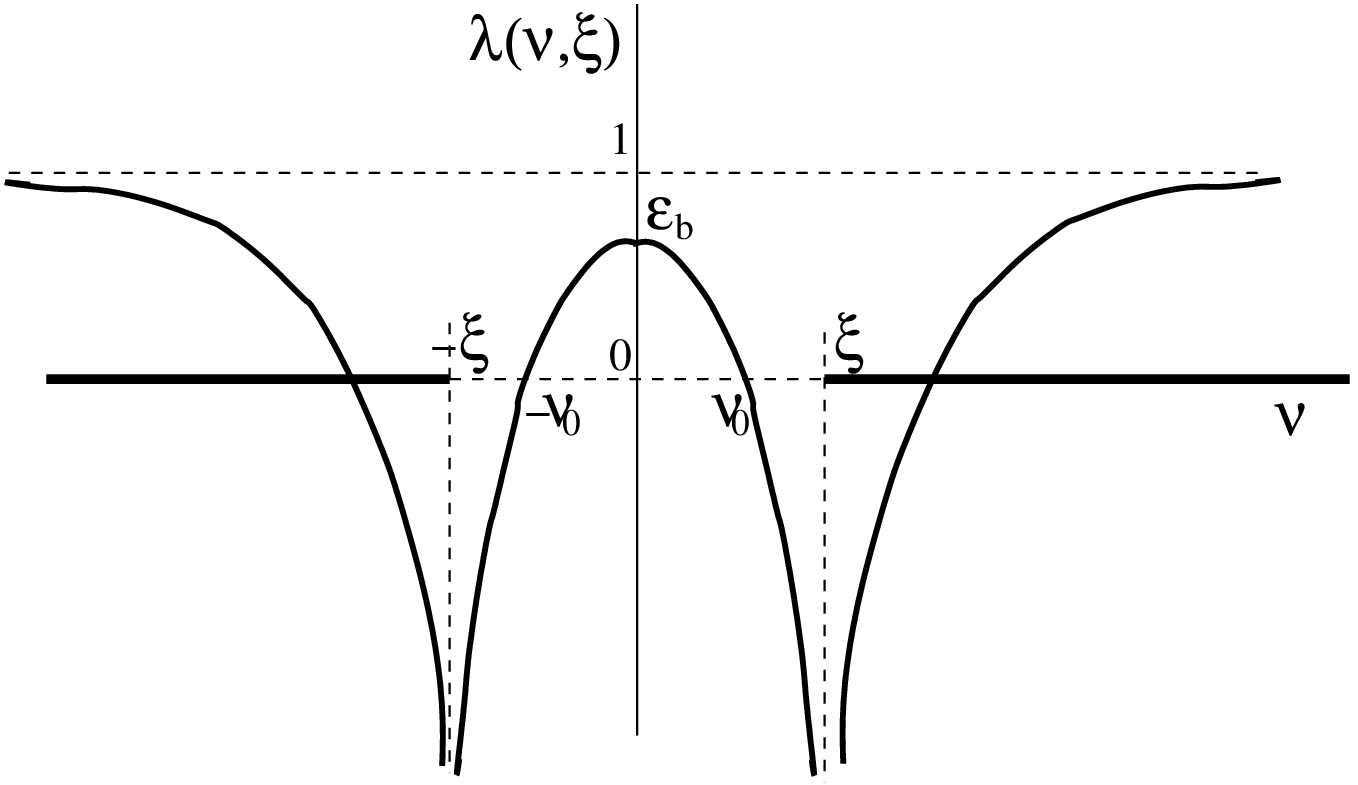}
\end{center}
\caption{The function $\lambda(\nu,\xi)$, real part of ${\cal L}^\pm(\nu,\xi)$, 
 as a function of $\nu$, for a given value of the
 frequency variable $\xi=\varphi(x)$. At the origin
  $\lambda(0,\xi)=\epsilon_{b}=1-\omega\,b$
 and at infinity $\lambda(\pm\infty)=1$. The
  zeroes at $\pm\nu_o(\xi)$ are also the zeroes
  of the dispersion function ${\cal L}(z,\xi)$. The latter is analytic
  in the complex plane cut along the semi-infinite intervals $\Re(z)=\nu\in
  ]-\infty,-\xi]\cup[+\xi,+\infty[$ indicated here by two thick lines.} 
\label{fig-lambda}
\end{figure}

A detailed description of the transformation of
Eq.~(\ref{eq-avintbis2}) into Eq.~(\ref{eq-cauchy2}) is given in the
Appendix~\ref{sec-whtocauchy}. The main idea is the separation of the 
integrations over $\tau'$ in  two ranges\,: $\tau'\in[-\infty,\tau]$ and
$\tau'\in[\tau,+\infty[$. Expressing then all the functions in
    Eq.~(\ref{eq-avintbis2}) in terms of their inverse
    Laplace transforms, simple algebra shows  that $j(\nu,\xi)$
        satisfies the singular integral equation \,: 
\begin{eqnarray}  
& & \lambda(\nu,\xi)j(\nu,\xi)= \omega\,b\,\xi\,
  k(\nu,\xi)\pvint_{-\infty}^{+\infty}\frac{j(\nu',\xi)}
  {\nu'-\nu}\,d\nu'\nonumber\\
& &  + \ \xi\left[p^\ast(\nu)t(\nu,\xi)
+ k(\nu,\xi)\pvint_{-\infty}^{+\infty}\frac{p^\ast(\nu')}{\nu'-\nu}\,d\nu'\right],
\label{eq-cauchy2}
\end{eqnarray}
where
\begin{equation}
  k(\nu,\xi)\equiv {1}/{2\nu},\,\,\nu>\xi, \quad k(\nu,\xi)\equiv
  0,\,\,0<\nu<\xi, 
  \label{def-knu}
\end{equation}
\begin{equation}  
  t(\nu,\xi)\equiv\pvint_\xi^\infty k(\nu',\xi) 
\left[\frac{1}{\nu' +\nu}+\frac{1}{\nu'
    -\nu}\right]\,d\nu,
\label{def-tnuxi}
\end{equation}
 and
\begin{equation}  
 \lambda(\nu,\xi)\equiv 1-\omega\,b\,\xi\,t(\nu,\xi).
\label{def-lambdanuxi}
\end{equation}
The symbol $\pvint$ means that the integral has to be taken in
principal part. Equation~(\ref{eq-cauchy2}) corresponds to Eq.~(18) in
\citet{hemsch72}. 
A sketch of $\lambda(\nu,\xi)$ for $\nu\in]-\infty,+\infty[$ is shown
in Fig.~\ref{fig-lambda}.  For a semi-infinite medium, $j(\nu,\xi)$
satisfies the same equation, but the range of
integration over $\nu'$ is limited to $[0,\infty[$. 

Equation~(\ref{eq-cauchy2}) has two singular integrals, one with
$j(\nu,\xi)$ and the other one with the frequency integrated function
$p^\ast(\nu)$. Its solution will be constructed in two
steps.  First it is assumed that $p^\ast(\nu)$ is a known
function. Equation~(\ref{eq-cauchy2}) represents then a singular
integral equation of the monochromatic type with a Cauchy kernel for
$j(\nu,\xi)$. Its solution, when integrated over $\xi$, provides a new
singular integral equation for $\bar\jmath(\nu)$, which is solved in
Section~\ref{sec-singular}.

The method of solution, which we apply to
Eq.~(\ref{eq-cauchy2}) and all the other singular integral equations
encountered in this article, was introduced by \citet{carleman22}. It
amounts to transform the singular integral equation into a boundary value
problem in the complex plane for the Hilbert transform of the unknown
function. Singular integral equations, Hilbert transforms, and their properties
are treated in detail in many books, such as \citet{musk53},
\citet{gakhov66}, \citet{carrier66}, \citet{pogo66},
\citet{williams71}, \citet{ablowitz97}.

We recall that the Hilbert transform of a function $f(t)$ is
the integral
\begin{equation}  
 {\cal H}(z)\equiv\frac{1}{2\ui\pi}\int_{C}\frac{f(t)}{t-z}\,dt,
\label{eq-defhilbert}
\end{equation}
where $z$ belongs to the complex plane and  $C$ is a given curve in
the complex plane. In the following we use the symbolic notation  
\begin{equation}  
  {\cal H}(z)=H_C[f(t)],
\label{eq-defhilbertsymb}
\end{equation}
where the subscript will indicate the integration curve. 
The function ${\cal H}(z)$ has very interesting properties. It
is analytic in the complex plane cut along the curve $C$, goes to
zero at infinity in the complex plane, is free of singularities, and
satisfies the Plemelj formulae. The latter state that for a point
$z_0$ on the cut, the limiting values ${\cal H}^+(z_0)$ and ${\cal H}^-(z_0)$ on
each side of it obey the relations
\begin{equation}  
  {\cal H}^+(z_0) - {\cal H}^-(z_0)=f(z_0),\quad
  {\cal H}^+(z_0) + {\cal H}^-(z_0)=
  \frac{1}{\ui\pi}\pvint_{C}\frac{f(t')}{t'-z_0}\,dt'. 
\label{eq-plemelj}
\end{equation}
In this article, the integration curve $C$ is always on the real line and
will be defined by its end-points. The Plemelj formulae will allows us
to transform Eq.~(\ref{eq-cauchy2}) into a boundary value problem in
the complex plane for the Hilbert transform of $j(\nu,\xi)$. The
latter is then retrieved with the first of the Plemelj formulae.

A function playing a critical role in the solution of
Eq.~(\ref{eq-cauchy2}) is the {\em dispersion function} constructed
with $\lambda(\nu,\xi)$. We now discuss its properties.
\section{The frequency-dependent dispersion function}
\label{sec-dispersionfreq}
The {\em dispersion function} is the analytic continuation of
$\lambda(\nu,\xi)$ in the complex plane. It is obtained by replacing in
Eq.~(\ref{def-lambdanuxi}) the real variable
$\nu$ by a complex variable $z$. The dispersion function is thus defined  by 
\begin{equation}  
{\cal L}(z,\xi)\equiv 1-\omega\,b\,\xi T(z,\xi),  
\label{eq-dispersion}
\end{equation}
where $z$ belongs to the complex plane and 
\begin{equation}  
  T(z,\xi)=\int_{{\cal C}(\xi)} k(\nu,\xi)\frac{1}{\nu -z}\,d\nu.
\label{eq-defT}
\end{equation}
$T(z,\xi)$ is actually the Fourier transform of the kernel $K_{\rm
  M}(\tau\xi)$, The notation ${\cal C}(\xi)$ stands for
$]-\infty,-\xi]\cup[+\xi,+\infty[$, that is the union of the two
intervals $]-\infty,-\xi]$ and $[+\xi,+\infty[$. The
 properties of ${\cal L}(z,\xi)$ are essentially the same as
those of the well studied monochromatic dispersion function
corresponding to $b=\xi=1$, henceforth denoted ${\cal L}_{\rm M}(z)$.
We summarized here the properties of ${\cal L}(z,\xi)$.

As shown by Eqs.~(\ref{eq-dispersion}) and (\ref{eq-defT})
$T(z,\xi)$, which is obviously a Hilbert transform, and ${\cal L}(z,\xi)$
are analytic in the complex plane cut along ${\cal C}(\xi)$.
At infinity,  $T(z,\xi) \to 0$ and ${\cal L}(z,\xi)\to 1$. For
$\nu\in{\cal C}(\xi)$, the Plemelj formulae lead to 
\begin{eqnarray}  
 T^+(\nu,\xi) - T^-(\nu,\xi) & = & 2\ui\pi\,k(\nu,\xi),\\
  T^+(\nu,\xi) + T^-(\nu,\xi)& = & 2\, t(\nu,\xi),\\ 
 {\cal L}^+(\nu,\xi) - {\cal L}^-(\nu,\xi) &=&
 -2\ui\pi\omega\,b\,\xi\,k(\nu,\xi),\\ 
{\cal L}^+(\nu,\xi) + {\cal L}^-(\nu,\xi)& =& 2\lambda(\nu,\xi).  
\label{dispplus}
\end{eqnarray}        
For $z\to 0$, 
\begin{equation}  
 {\cal L}(z,\xi)\simeq \epsilon_{b} - (1-\epsilon_{b})\frac{{z}^2}{3\xi ^2}, 
\label{eq-lzero}
\end{equation}
hence
\begin{equation}  
{\cal L}(0,\xi)=\epsilon_{b}.  
\label{eq-lzero2}
\end{equation}
Here,
\begin{equation}  
  \epsilon_{b}= 1-\omega\,b=1-(1-\epsilon)\,b.
\label{def-epsb}
\end{equation}
The parameter $\epsilon_{b}$ is  positive and smaller than unity since
$b\in[0,1]$.

Another important property of ${\cal L}(z,\xi)$ is that it  has a pair
of real zeroes in the interval $[-\xi,+\xi]$ coinciding with
the zeroes of $\lambda(\nu,\xi)$ at $\pm\nu_0(\xi)$ (see
Fig.~\ref{fig-lambda}). The dispersion function has no other zero.
The two zeroes and the cut along ${\cal C}(\xi)$ are known as the
{\em spectrum} of $\nu$ or as the {\em discrete} and
{\em continuous eigenvalues} of the eigenfunctions $\Phi(k,x,\mu)$
\citep{cormick73}. We refer here to the interval $[-\xi,+\xi]$
containing the discrete eigenvalues as the {\em discrete spectrum} of
${\cal L}(\nu,\xi)$ and to ${\cal
  C}(\xi)=]-\infty,-\xi]\cup[+\xi,+\infty[$ as the {\em continuous spectrum}.  

The integration over $\nu$ in Eq.~(\ref{eq-defT}) is easily carried
out and leads to the explicit expression
\begin{equation}  
  {\cal L}(z,\xi)= 1- \omega\,b\,\xi\frac{1}{2z}\ln
\frac{\xi + z}{\xi-z},
\label{eq-dispersion2}
\end{equation}
which shows that ${\cal L}(z,\xi)$ is actually a function of
the complex variable  
\begin{equation}  
 \tilde z=z/\xi. 
\label{def-newvar}
\end{equation} 
and that we have ${\cal L}(z,\xi)=\tilde{\cal L}(\tilde z)$ with
\begin{equation}  
  \tilde{\cal L}(\tilde z)= 1- (1-\epsilon_{b})\frac{1}{2\tilde z}
\ln\frac{1 + \tilde z}{1-\tilde z}. 
\label{eq-dispersion3}
\end{equation}
The zeroes of $\tilde{\cal L}(z)$ are located at $\pm u_0$, with $u_0$
solution of $\tilde{\cal L}(u_0)=0$ and those of ${\cal L}(z,\xi)$ at
$\nu_0(\xi)=\xi u_0$. The inverse relation giving the
value of $\xi$ for a given position $\nu$ of the positive zero is
$\xi_0(\nu)=\nu/u_0$.

The partial derivatives of the dispersion function at
$\pm\nu_0(\xi)$ play an important role in the
calculation of $j(\nu,\xi)$ and ${\bar \jmath}(\nu)$. At these points
${\cal L}(z,\xi)=\lambda(\nu,\xi)$, with $\nu=\Re(z)$. Using
Eq.~(\ref{eq-dispersion3}), one readily obtains 
\begin{equation}  
 \frac{\partial \lambda(\nu,\xi)}{\partial
   \nu}|_{\pm \nu_{0}}=\pm\frac{1}{\xi u_0}\left[\frac{\epsilon_{b}
      -u_0 ^2}{1 -u_0 ^2}\right],  
\label{eq-derivativenu}
\end{equation}
and
\begin{equation}  
 \frac{\partial \lambda(\nu,\xi)}{\partial
   \xi}|_{\pm \nu_{0}}=\mp\frac{u_0}{\nu_0}\left[\frac{\epsilon_{b}
      -u_0 ^2}{1 -u_0 ^2}\right].  
\label{eq-derivativexi}
\end{equation}
The derivatives with respect to $\nu$ and $\xi$ have opposite signs,
the derivative with respect to $\nu$ being negative and the other one
positive. Their ratio is $-1/u_0$.

\section{The infinite medium}
\label{sec-singular}
We now apply the Carleman method to solve the singular integral
equation for $j(\nu,\xi)$ written in Eq.~(\ref{eq-cauchy2}). In this
equation, the coefficient $\lambda(\nu,\xi)$, which appears in the
left-hand side, has zeroes in the interval
$[-\xi,+\xi]$ at $\pm\nu_0(\xi)$. For $\nu\ne\pm\nu_0(\xi)$,
$j(\nu,\xi)$ is given by the right-hand side
divided by $\lambda(\nu,\xi)$. For $\nu=\pm\nu_0(\xi)$, the right-hand
side of Eq.~(\ref{eq-cauchy2}) has no reason to be
zero. Equation~(\ref{eq-cauchy2})  
 can be satisfied for $\nu=\pm\nu_0(\xi)$ by assuming that 
$j(\nu,\xi)=f(\nu,\xi)[\delta(\nu-\nu_0(\xi)) +
  \delta(\nu+\nu_0(\xi))]$, where $f(\nu,\xi)$ is at this stage an
arbitrary function. It is indeed a property of the Dirac delta
function that $\int x\delta(x)\,dx=0$. The solution of
Eq.~(\ref{eq-cauchy2}) valid for all values of $\nu$ is given in
Eq.~(\ref{eq-solcenter2}). The same argument is found in the
construction of the singular eigenfunctions of the \citet{case60}
expansion method.

Hilbert transforms are usually introduced for
H\"older continuous functions \citep[e.g.,][]{musk53}. This means that
one should treat separately the intervals $]-\xi,+\xi[$ and
${\cal C}(\xi)$. This is done  in Section~\ref{sec-half}
for the semi-infinite medium. Hilbert transforms can however still be defined
for a Dirac distribution \citep{pandey96} and the Plemelj formulae
remain applicable. Following \citet{hemsch72}, the Hilbert
transforms are defined in this Section over the full real axis. 
Using the notation defined in
Eq.~(\ref{eq-defhilbertsymb} we introduce\,: 
\begin{equation}  
  {N}_\infty(z,\xi)\equiv H_{-\infty}^{+\infty}[j(\nu,\xi)],\quad {\bar
    N}_\infty(z)\equiv H_{-\infty}^{+\infty}[\bar\jmath(\nu)],
\label{hilbert}
\end{equation}
  and
 \begin{equation}
 N_{\rm Q}^\infty(z)\equiv H_{-\infty}^{+\infty}[q^\ast(\nu)],\quad
 P_\infty(z)\equiv H_{-\infty}^{+\infty}[p^\ast(\nu)].
 \label{eq-defnq}
 \end{equation}
The Hilbert transform of $q^\ast(\nu)$ has all the required properties when the
integral of $Q^\ast(\tau)$ over $\tau$ has a finite value. By
construction, these four Hilbert transforms are   
analytic in the complex plane cut along the real axis and go to zero
as $z$ goes to infinity.

In Section~\ref{sec-singular1} we determine ${N}_\infty(z,\xi)$ in
terms of ${\bar N}_\infty(z)$ and in Section~\ref{sec-singular2}
${\bar N}_\infty(z)$ by an integration over $\xi$ of
${N}_\infty(z,\xi)$. The functions $j(\nu,\xi)$ and
$\bar\jmath(\nu)$ are then given by the 
jumps of ${N}_\infty(z,\xi)$ and  ${\bar N}_\infty(z)$ across the real
axis (see Eq.~(\ref{eq-plemelj})). Explicit expressions of $j(\nu,\xi)$ and
$\bar\jmath(\nu)$ are given in Section~\ref{sec-singular3} 
for the Green's function.

\subsection{The frequency dependent Hilbert transform $N_\infty(z,\xi)$}
\label{sec-singular1}

Making use of the expression of ${\cal L}(z,\xi)$ in
Eq.~(\ref{eq-dispersion}), of the Plemelj formulae for $T(z,\xi)$ and
${\cal L}(z,\xi)$, and also of the Hilbert transforms in
Eqs.~(\ref{hilbert}) and (\ref{eq-defnq}), we can rewrite
Eq.~(\ref{eq-cauchy2}) as 
\begin{eqnarray}  
& & {\cal L}^+(\nu,\xi)N^+_\infty(\nu,\xi)- {\cal
    L}^-(\nu,\xi)N^-{_\infty}(\nu,\xi)= \nonumber\\
& & \frac{1}{\omega\,b}\Bigl\{[1-{\cal L}^+(\nu,\xi)]P_{\infty}^+(\nu)-
    [1-{\cal L}^-(\nu,\xi)]P_{\infty}^-(\nu)\Bigr\}.
\label{eq-cauchy4}
\end{eqnarray}
We stress that in this Section devoted to frequency dependent
functions, we necessarily have $b\ne 0$.  To solve
Eq.~(\ref{eq-cauchy4}) for $N_\infty(z,\xi)$, we use a standard method
based on the Liouville theorem.  Equation~(\ref{eq-cauchy4}) is
equivalent to
\begin{equation}  
  F^+(\nu,\xi)- F^-(\nu,\xi)= 0, \quad \nu\in]-\infty,+\infty[,
  \label{eq-defF}
  \end{equation}
where
\begin{equation}  
 F(z,\xi)\equiv{\cal L}(z,\xi)N_\infty(z,\xi) 
- \ \frac{1}{\omega\,b}[1-{\cal L}(z,\xi)]P_{\infty}(z).
\label{def-fplus}
\end{equation}
By construction, $F(z,\xi)$ is analytic in the complex plane cut along
the full real axis and, according to
  Eq.~(\ref{eq-defF}), the jump of $F(z,\xi)$ across this axis is
  zero.  Hence $F(z,\xi)$ is an entire function, which can be
  determined by its behavior at infinity, by application of the
  Liouville theorem (e.g. Carrier et al. 1966). For $z\to\infty$,
  $N_\infty(z,\xi)$ and $P_{\infty}(z)$ tend to 0 and
  ${\cal L}(z,\xi)\to 1$. Hence,  $F(z,\xi)\to 0$ as $z\to \infty$
  and by the Liouville theorem $F(z,\xi)$ is zero for all $z$. We thus obtain
\begin{eqnarray}  
  & & {\cal L}(z,\xi) {N}_\infty(z,\xi)\nonumber\\
  & & =\ \frac{1}{\omega\,b}[1-{\cal
    L}(z,\xi)]P_\infty(z)=\frac{1}{\omega\,b}[1-{\cal
    L}(z,\xi)][\omega(1-b)\bar N_\infty(z) + N^\infty_{\rm Q}(z)]. 
\label{eq-hemsch}
\end{eqnarray}
 Since $ {\cal L}(z,\xi)$ has zeroes at $\pm
 \nu_0(\xi)\in[-\xi,+\xi]$, Eq.~(\ref{eq-hemsch}) cannot be
 straightforwardly divided by ${\cal L}(z,\xi)$. We must distinguish
 between the regions $z\in[-\xi,+\xi]$ and $z$ outside this interval. 

For $z\notin[-\xi,+\xi]$, we can divide
Eq.~(\ref{eq-hemsch}) by ${\cal L}(z,\xi)$. Using the expression
of ${\cal L}(z,\xi)=1-\omega\,b\,\xi T(\nu,\xi)$ (see
Eq.~(\ref{eq-dispersion})), we obtain 
\begin{equation}  
  {N}_\infty(z,\xi)=\xi\frac{T(z,\xi)}{{\cal L}(z,\xi)}
  [\omega(1-b)\bar N_\infty(z) + N^\infty_{\rm Q}(z)].
\label{eq-relation}
\end{equation}

In the interval $[-\xi, +\xi]$, according to  Eq.~(\ref{def-knu}),
$k(\nu,\xi)=0$, hence Eq.~(\ref{eq-cauchy2}) reduces to  
\begin{equation} 
 \lambda(\nu,\xi)j(\nu,\xi)= \xi\,p^\ast(\nu)\,t(\nu,\xi). 
\label{eq-solcenter}
\end{equation}
As explained at the beginning of Section~(\ref{sec-singular}), because
of the zeroes of $\lambda(\nu,\xi)$ at $\pm \nu_0(\xi)$, the 
solution of Eq.~(\ref{eq-solcenter}) has the general form
\begin{equation} 
j(\nu,\xi)=f(\nu,\xi)[\delta(\nu-\nu_0(\xi)) +
    \delta(\nu+\nu_0(\xi))]
 + \ p^\ast(\nu){P}\frac{\xi\,t(\nu,\xi)}{\lambda(\nu,\xi)},  
\label{eq-solcenter2}
\end{equation}
where $\delta(\nu\pm\nu_0(\xi))$ are Dirac distributions at
$\pm\nu_0(\xi)$ and $P$ stands for Cauchy principal part. Using
$\int x\delta(x)\,dx=0$, we can check that the solution in
Eq.~(\ref{eq-solcenter2}) satisfies Eq.~(\ref{eq-solcenter}) for all
values of $\nu$. 

The function
$f(\nu,\xi)$ can be determined by considering the limit $z\to \pm
\nu_0(\xi)$ in Eq.~(\ref{eq-hemsch}). In the right-hand side we can
set ${\cal L}(z,\xi)=0$. In the left-hand side, the only terms giving
a non-zero contribution come from the Dirac distributions. We then
Taylor expand ${\cal L}(z,\xi)$ around its zeroes at $\pm \nu_0(\xi)$,
that is assume ${\cal L}(z,\xi)\simeq (z\pm\nu_0)\partial_\nu{\cal
  L}(\pm\nu_0,\xi)$, where $\partial_\nu{\cal
  L}(\pm\nu_0,\xi)=\partial_\nu \lambda(\pm\nu_0,\xi)$ is the partial
derivative of $\lambda(\nu,\xi)$ with respect to $\nu$ taken at
$\pm\nu_0(\xi)$. Equation~(\ref{eq-hemsch}) thus leads to
\begin{equation}
f(\pm\nu_0,\xi)\partial_\nu\lambda(\pm\nu_0,\xi)=-\frac{2\ui\pi}{\omega\,b}
\Big\langle P_\infty(\pm\nu_0)\Big\rangle, 
\label{eq-fpmnuzero2}
\end{equation}
where
\begin{equation}
 \Big\langle{P}_\infty(\pm\nu_0)\Big\rangle\equiv \frac{1}{2}\left[{
     P}_\infty^+(\pm\nu_0)+
   {P}_\infty^-(\pm\nu_0)\right]=\frac{1}{2\ui\pi}
 \pvint_{-\infty}^{+\infty}\frac{p^\ast(\nu')}{\nu'\mp \nu_0}\,d\nu'.
\label{eq-defangle} 
\end{equation} 
We recall that ${P}_\infty(z)$, the Hilbert transform of $p^\ast(\nu)$
is analytic in the complex plane cut along the real axis. The
derivative $\partial_\nu\lambda(\pm\nu_0,\xi)$ is given in
Eq.~(\ref{eq-derivativenu}).  One can check that $f(\pm\nu_0,\xi)$ is
positive.

The expressions that we have obtained for $N_\infty(z,\xi)$ and
$f(\pm\nu_0,\xi)$ are still containing the unknown function $\bar
{N}_\infty(z)$. We now proceed to the determination of the latter.

\subsection{The frequency integrated Hilbert transform $\bar N_\infty(z)$}
\label{sec-singular2}
Knowing how to properly define $N_\infty(z,\xi)$ for all $z$, we
can use Eq.~(\ref{eq-relation}) to determine $\bar
N_\infty(z)$. Integrating over $\xi$, we obtain  
\begin{equation}  
  \Omega(z){\bar N}_\infty(z)=N_{\rm Q}^\infty(z)\int_0^{\varphi_0}W(\xi)\xi
  \frac{T(z,\xi)}{{\cal L}(z,\xi)}\,d\xi,
  \label{eq-nbareq}
\end{equation}
where $W(\xi)$ is the weight function defined in Eq.~(\ref{def-wxi}), and
\begin{equation}  
  \Omega(z)\equiv 1-\omega(1-b)\int_0^{\varphi_0}W(\xi)\xi
  \frac{T(z,\xi)}{{\cal L}(z,\xi)}\,d\xi.
  \label{def-omega2}
\end{equation}
The function $\Omega(z)$ plays for ${\bar N}_\infty(z)$ the role
played by the dispersion function ${\cal L}(z,\xi)$ for
${N}_\infty(z,\xi)$. It will be referred to as the {\em generalized
  dispersion function}. Equation~(\ref{def-omega2}) shows that
$\Omega(z)$ is analytic in $\cset\ /\rset$.\footnote{The notation
  $\cset\ /\rset$ stands for the complex plane cut along the full real
  axis} It is shown in 
Appendix~\ref{sec-omega1} that $\Omega(z)$ has no zero and in
Appendix~\ref{sec-omega2} that it goes to the complete frequency
dispersion function for $b=0$. The latter is denoted ${\cal L}_{\rm
  R}(z)$, the subscript R standing for complete  frequency 
redistribution. For $b=1$, that is in the monochromatic limit, one simply has 
$\Omega(z)=1$.  Other properties of $\Omega(z)$ are described in
Appendix~\ref{sec-omega1}.

Since $\Omega(z)$ has no zero, we can divide Eq.~(\ref{eq-nbareq})
by $\Omega(z)$. Making use of Eq.~(\ref{def-omega2}), we obtain for
$\bar N_{\infty}(z)$ an explicit expression, analytic in
$\cset\ /\rset$, which may be written as
\begin{equation}  
\bar N_\infty(z)=\frac{1}{\omega(1-b)}N_{\rm
  Q}^\infty(z)\left[\frac{1}{\Omega(z)} -1\right].  
\label{eq-nbarinfini}
\end{equation}

We can now obtain closed form expressions for $N_\infty(z,\xi)$ and
$f(\pm\nu_0,\xi)$. Inserting Eq.~(\ref{eq-nbarinfini}) into
Eqs.~(\ref{eq-relation}) and (\ref{eq-fpmnuzero2}), we obtain 
\begin{equation}  
  N_\infty(z,\xi)=\frac{1}{\omega\,b}N_{\rm
    Q}^\infty(z)\left[\frac{1}{{\cal L}(z,\xi)} -1\right]\frac{1}{\Omega(z)},
\label{eq-nzedxi3}
\end{equation}
and
\begin{equation}
f(\pm\nu_0,\xi)\partial_\nu\lambda(\pm\nu_0,\xi)=-\frac{2\ui\pi}{\omega\,b}
\Big\langle{\bar N}_{\rm Q}(\pm\nu_0)\frac{1}{\Omega(\pm\nu_0)}\Big\rangle.
\label{eq-fpmnuzero3}
\end{equation}
The expression given for  $\bar N_\infty(z)$ holds only when
$b\ne 1$. For $b=0$ it provides the complete frequency redistribution
limit. The expressions of $N_\infty(z,\xi)$ and $f(\pm\nu_0,\xi)$
hold only for $b\ne 0$. For $b=1$, they provide the monochromatic limit.

\subsection{Explicit expressions of $j(\nu,\xi)$ and
  $\bar\jmath(\nu)$ for the Green's function}
\label{sec-singular3}

We give here explicit expressions of $j(\nu,\xi)$ and
$\bar\jmath(\nu)$ obtained by setting
\begin{equation}  
 Q^\ast(\tau)=\delta(\tau). 
\label{eq-primgreen}
\end{equation}
For this special primary source, $S(\tau,\xi)$ is the infinite
medium Green's function and $(1-\epsilon){\cal J}(\tau,\xi)$, its
regular part, often referred to as the resolvent function.
Replacing  $Q^\ast(\tau)$ by $\delta(\tau)$ in  the integral equation for 
${\cal J}(\tau,\xi)$ (see Eq.~(\ref{eq-avintbis2})), we observe that
the contribution from the primary source term  is now given by 
$\xi K_{\rm M}(|\tau|\xi)$, the inverse Laplace transform of which is
\begin{equation}  
  \xi\,k(\nu,\xi)=-\frac{1}{2\ui\pi\omega\,b}[{\cal
      L}^+(\nu,\xi)-{\cal L}^-(\nu,\xi)].
\label{eq-dispmoins2}
\end{equation}
Equation~(\ref{eq-cauchy2}) still provides the integral equation for
$j(\nu,\xi)$. It suffices to add  $\xi\,k(\nu,\xi)$ to the right-hand
side, while simultaneously setting $q^\ast(\nu)=0$. 

Proceeding as described in Section~\ref{sec-singular1}, we obtain the
relation
\begin{equation}  
  {\cal L}(z,\xi) {N}_\infty(z,\xi)=[1-{\cal
      L}(z,\xi)]\left[\frac{1-b}{b}{\bar N}_\infty(z) +
    \frac{1}{\omega\,b2\ui\pi}\right]. 
\label{eq-hemschgreen}
\end{equation}
Comparing it with Eq.~(\ref{eq-hemsch}), we see that $N_{\rm Q}^\infty(z)$
has been replaced  by $1/(2\ui\pi)$. It suffices to carry out this change in
Eqs.~(\ref{eq-nbarinfini}), (\ref{eq-nzedxi3}) and
(\ref{eq-fpmnuzero3}) to obtain
$\bar N_\infty(z)$, $N_\infty(z,\xi)$ and $f(\pm\nu_0,\xi)$. 

The function $\bar\jmath(\nu)$ is given by the jump of $\bar
N_\infty(z)$ across the real axis. Taking advantage of the fact that
$\Omega ^+(\nu)$ and $\Omega ^-(\nu)$ are complex conjugate, we can write
\begin{equation}  
\bar \jmath (\nu)=\frac{1}{\omega\pi(1-b)}\frac{\Im[-\Omega ^+(\nu)]}
{[\Re[\Omega ^+(\nu)]] ^2+ [\Im[\Omega ^+(\nu)]]^2}. 
\label{eq-jmean}
\end{equation}
When $\Im[-\Omega ^+(\nu)]$ is replaced by the explicit expressions given by
Eqs.~(\ref{eq-omegapm1}), (\ref{eq-omegapm2})
(\ref{eq-omegapm3}), the factor $\omega(1-b)$ disappears, but there
remains a dependence on $b$, which appears in the functions defined in
Eqs.~(\ref{def-grandC}) and (\ref{def-productpm}).
 The complete frequency redistribution limit $b=0$ is
recovered when $\Omega(z)$ is replaced by the complete frequency
redistribution dispersion function ${\cal L}_{\rm R}(z)$. In this case
$\bar\jmath(\nu)$ becomes
\begin{equation}  
  \bar \jmath_{\rm R}(\nu)=\frac{k_{\rm R}(\nu)}{{\cal
      L}^+_{\rm R}(\nu){\cal L}^-_{\rm R}(\nu)},
\label{eq-jbarnulimcr}
\end{equation}
where $k_{\rm R}(\nu)$ is the inverse Laplace transform of the kernel
$K_{\rm R}(\tau)$ (see Eqs.~(\ref{eq-kernelc2})).

The case of $j(\nu,\xi)$ is slightly more complex since we have to
distinguish between the discrete part of the spectrum corresponding to
$\nu \in[-\xi,+\xi]$ and the continuous part for $\nu\in{\cal C}(\xi)$.
As shown by Eq.~(\ref{eq-solcenter2}), the discrete part contains the
two Dirac distributions at $\pm\nu_0(\xi)$,  
corresponding to the zeroes of $\lambda(\nu,\xi)$, plus another term
taken in principal value because it has $1/\lambda(\nu,\xi)$ in the denominator.
It is convenient to write the discrete part, denoted
$j_{\rm d}(\nu,\xi)$, with d standing for discrete, as  
\begin{equation}  
  j_{\rm d}(\nu,\xi)= j_{\rm d}^{(0)}(\nu,\xi) + j_{\rm d}^{(1)}(\nu,\xi), 
\label{eq-jnud}
\end{equation}
where the superscript 0 correspond to the Dirac distributions. The 
term $j_{\rm d}^{(1)}(\nu,\xi)$ is given by  
the jump of $N_\infty(z,\xi)$ across the real axis for $\nu\in[-\xi,+\xi]$.
Equations~(\ref{eq-nzedxi3}) and (\ref{eq-fpmnuzero3}) in which
$N_{\rm Q}^\infty(z)$ is replaced by $1/(2\ui\pi)$ lead to
\begin{equation} 
j_{\rm d}^{(1)}(\nu,\xi)=
 \frac{\xi}{\pi}\,\frac{\Im[-\Omega^+(\nu)]} {\Omega
  ^+(\nu)\Omega^-(\nu)}P\frac{t(\nu,\xi)}{\lambda(\nu,\xi)},
\label{eq-jnudiscrete2}
\end{equation}
\begin{equation}  
 j_{\rm d}^{(0)}(\nu,\xi) = f(\nu,\xi)[\delta(\nu-\nu_0(\xi)) +
   \delta(\nu+\nu_0(\xi))], 
\label{eq-jnud0}
\end{equation}
with
\begin{equation}  
  f(\nu,\xi)=-\ \frac{1}{\omega\,b}\,\frac{1}{\partial{\lambda}_\nu(\nu,\xi)}\, 
\Big\langle\frac{1}{\Omega ^\pm(\nu_0)}\Big\rangle.
\label{eq-deffpole}
\end{equation}
The angle bracket is defined in Eq.~(\ref{eq-defangle}). One can check
that $j_{\rm d}(\nu,\xi)$ 
is positive. In the monochromatic limit 
$j_{\rm d}(\nu,\xi)$ is reduced to $j_{\rm d}^{(0)}(\nu,\xi)$ since 
$\Omega(z)=1$ when  $b=1$ (see Eq.~(\ref{def-omega2})).
 
In the continuous part of the spectrum, $j(\nu,\xi)$ is given by the
jump of $N_\infty(z,\xi)$ across the real axis for $\nu\in{\cal
  C}(\xi)$.  Combining Eq.~(\ref{eq-nzedxi3}) and the Plemelj
formulae for ${\cal L}(z,\xi)$, we can write  $j(\nu,\xi)$ as
\begin{equation}  
j_{\rm c}(\nu,\xi)=\xi\frac{k(\nu,\xi)}{{\cal
      L}^+(\nu,\xi){\cal L}^-(\nu,\xi)}
+ \ \frac{1}{2\ui\pi\omega\,b}\left[\frac{(1-\Omega(z))(1-{\cal
        L}(z,\xi))}{\Omega(z){\cal L}(z,\xi)}\right]_{\rm jmp},
\label{eq-jxicont2}
\end{equation}
where the subscript ${\rm jmp}$ stands for jump across the real axis and
the subscript c for continuous. In the monochromatic limit
the right-side is reduced to its first term since $\Omega(z)=1$ for $b=1$. 
There is no point in considering the complete frequency limit $b=0$ 
since we are dealing here with a frequency dependent function.

We want to stress here the roles of the zero of ${\cal L}(z,\xi)$ at
$\nu_0(\xi)$. First it gives a Dirac distribution in $j_{\rm
  d}(\nu,\xi)$, hence an exponential term $\ue ^{-\nu_0(\xi)\tau}$ in
the source function $S(\tau,\xi)$ as shown in
Eq.~(\ref{eq-sourcegene}). Additional effects from the zeroes of
${\cal L}(z,\xi)$ appear in the functions $f(\nu_0,\xi)$, $j_{\rm
  d}^{(1)}(\nu,\xi)$, and also in $\bar\jmath(\nu)$ for
$\nu\in[0,\xi]$, via the derivatives of ${\cal L}(z,\xi)$ with respect
to $\nu$ or $\xi$ taken at $\nu_0(\xi)$.  A derivative with respect to
$\nu$ appears explicitly in $f(\nu_0,\xi)$ (see
Eq.~(\ref{eq-deffpole})).  The derivative with respect to $\xi$ sneaks
in for $\nu\in[0,\xi]$, through the imaginary part of
$\Omega^\pm(\nu)$, which, as shown in Eq.~(\ref{eq-omegapm3}), has a
term of the form
\begin{equation}
-\pi\omega(1-b)\,W(\xi_0(\nu))\xi_0(\nu)
  \frac{t(\nu,\xi_0(\nu))}{\partial_\xi{\cal L}(\nu,\xi_0(\nu))}.  
\label{eq-meanpole1} 
\end{equation}
The zeroes of ${\cal L}(z,\xi)$ have thus a very important role for
all values of $b$, except for $b=0$. In this limit, the
dispersion function ${\cal L}_{\rm R}(z)$ is free of zeroes. 

\subsection{The source function in the physical space}
\label{sec-jnujnu}
In the physical space the source function is given  by
\begin{equation}  
 S(\tau,\xi)=\omega[b\,{\cal J}(\tau,\xi) + (1-b)\bar {\cal J}(\tau)]
+ Q^\ast(\tau).
\label{def-green}
\end{equation}
We limit the discussion to positive values of $\tau$, for which we have  
\begin{equation}  
{\cal J}(\tau,\xi)=\int_0^\infty j(\nu,\xi)\,\ue
^{-\nu\tau}\,d\nu,\quad  \bar{\cal J}(\tau)=\int_0^\infty \bar \jmath(\nu)\,\ue
^{-\nu\tau}\,d\nu. 
\label{eq-inverseplus}
\end{equation}
We recall that $\xi=\varphi(x)$, with $\varphi(x)$ the
absorption line profile. The analysis of $j(\nu,\xi)$ and
$\bar\jmath(\nu)$ carried out in the preceding sections shows that  we
can write the source function as 
\begin{eqnarray}  
& &   S(\tau,\xi)=\omega\left[b f(\nu_0(\xi),\xi)\ue ^{-\nu_0(\xi)\tau} +
    \int_0^\xi[b j_{\rm d}^{(1)}(\nu,\xi) + (1-b)\bar\jmath(\nu)]\ue
      ^{-\nu\tau}\, d\nu\right.\nonumber\\
    & & \left.\ +  \int^\infty_\xi  [b j_{\rm c}(\nu,\xi) +
      (1-b)\bar\jmath(\nu)]\ue ^{-\nu\tau}\, d\nu\right] +  Q^\ast(\tau),
\label{eq-sourcegene}
\end{eqnarray}
where the subscripts d and c stand for discrete and continuous.

In the monochromatic limit corresponding to $b=1$,
Eq.~(\ref{eq-sourcegene}) becomes 
\begin{equation}  
   S_{\rm M}(\tau,\xi)=\omega\left[f(\nu_0(\xi),\xi)\ue ^{-\nu_0(\xi)\tau} +
     \int^\infty_\xi j_{\rm c}(\nu,\xi)\ue ^{-\nu\tau}\,d\nu\right] +
   Q^\ast(\tau).  
\label{eq-sourcemono}
\end{equation}
There is no contribution from the term $j_{\rm d}^{(1)}(\nu,\xi)$, 
which is zero  as shown by Eq.~(\ref{eq-jnudiscrete2}) because
$\Omega(z)=1$. For the same reason, $j_c(\nu,\xi)$ is reduced to the
first term in the right-hand side of Eq.~(\ref{eq-jxicont2}). When
$\xi$ is set to one, one recovers the usual monochromatic source
function.  For $\tau\to\infty$, $S_{\rm M}(\tau,\xi)$ decreases 
exponentially as $\ue ^{-\nu_0(\xi)\tau}$. The integral term decreases
faster since its lower limit $\xi$ is larger than $\nu_0(\xi)$. 

In the complete frequency redistribution limit, that is for $b=0$, we
readily obtain
\begin{equation}  
   S_{\rm R}(\tau,\xi)=\omega\int_0^\infty \bar\jmath(\nu)\,\ue
   ^{-\nu\tau}\,d\nu  + Q^\ast(\tau).  
\label{eq-sourcecplt}
\end{equation}
It is quite obvious that $S_{\rm R}(\tau,\xi)$ is independent of
$\xi$. For $\tau\to\infty$, $S_{\rm R}(\tau,\xi)$ decreases algebraically as a
power law $\tau ^{-\alpha}$ with $\alpha>0$ depending on the behavior of
$\varphi(x)$ for $x$ large. For a Doppler (Gaussian) profile,
$\alpha=1$ (there is also a logarithmic correction). For a Voigt
profile $\alpha=1/2$  \citep[see e.g.,][p.~159]{ivanov73}.

Equation~(\ref{eq-sourcegene}) shows that as soon as $b$ becomes
smaller than one, there will be a competition between the exponential
term and the first integral in which $\nu$, belonging to the interval
$[0,\xi]$, can take values smaller than $\nu_0(\xi)$.  In this interval
$\Omega ^\pm(\nu)$ depends on a derivative at $\nu_0(\xi)$ of the
monochromatic dispersion function (see Eq.~(\ref{eq-omegapm3})), hence
the functions  $\bar\jmath(\nu)$ and $j_{\rm d}^{(1)}(\nu)$ depend
also on this derivative (see Eqs.~(\ref{eq-jmean}) and
(\ref{eq-jnudiscrete2})). The zero of the  
dispersion function at $\nu_0(\xi)$ has thus several roles. It appears
directly in the pure  exponential term  but also indirectly through
derivatives of the dispersion function.   

A very precise asymptotic expansion for $\nu\to 0$ or numerical calculations
are needed to describe in some detail the transition between the
monochromatic and the complete frequency redistribution limits, that is
between an exponential and an algebraic large $\tau$ asymptotic
behavior. This point is further discussed in Section~\ref{sec-conclusion}.

\section{The semi-infinite medium}
\label{sec-half}

We now consider a semi-infinite plane-parallel medium, where the space
variable $\tau$ varies from 0 at the surface to $\infty$ in the
interior. We also assume that there is no incident radiation field at
$\tau=0$. An incident field can always be transformed into an interior
source by separating the contribution of the incident field from that
of the diffuse field created by the scattering. The functions ${\cal
  J}(\tau,\xi)$ and $\bar{\cal J}(\tau)$ for $\tau\ge 0$ are defined
as shown in Eqs.~(\ref{eq-inverse1}) to (\ref{def-barjnu}). Now $\nu$
has only positive values.

We first determine $j(\nu,\xi)$ (Section~\ref{sec-halffreq}) and then
$\bar\jmath(\nu)$ by integration over $\xi$
(Sections~\ref{sec-semiintegral} and \ref{sec-fredholm}). Exact
solutions of half-space problems always depend on a half-space
auxiliary function derived from the dispersion function. We will
introduce two half-space auxiliary functions, namely 
$X(z,\xi)$ constructed with ${\cal L}(z,\xi)$ and $\bar X(z)$,
constructed with $\Omega(z)$. We use a standard construction method
based on the solution of a homogeneous Riemann--Hilbert problem
\citep[e.g.,][]{ablowitz97}. Finally we
show in Section~\ref{sec-emergent} how to calculate the emergent
intensity.

\subsection{The frequency dependent function $j(\nu,\xi)$}
\label{sec-halffreq}

The singular integral equation for $j(\nu,\xi)$ is
given by Eq.~(\ref{eq-cauchy2}), with the domain of definition of $\nu$
restricted to $[0,\infty[$. This means that we now have to solve
\begin{eqnarray}  
& & \lambda(\nu,\xi)j(\nu,\xi)= \omega\,b\,\xi\,
  k(\nu,\xi)\pvint_0^{\infty}\frac{j(\nu',\xi)}{\nu'-\nu}\,d\nu'\nonumber\\
& & + \ \xi\left[p^\ast(\nu)\,t(\nu,\xi) + k(\nu,\xi)\pvint_0^{\infty}
\frac{p^\ast(\nu')}{\nu'-\nu}\,d\nu'\right],
\label{eq-cauchyhalf}
\end{eqnarray}
where $p^\ast(\nu)$ has been defined in Eq.~(\ref{eq-defpstar1}). For
the full-space case, we have introduced $N_\infty(z,\xi)$, the Hilbert
transform of $j(\nu,\xi)$ over the full real axis.  Here, we consider
separately the interval $\nu\in[0,\xi]$, where $j(\nu,\xi)$ has a
distributional solution, and the interval $\nu\in[\xi,\infty[$, where
an auxiliary function $X(z,\xi)$ is needed to determine $j(\nu,\xi)$.

\subsubsection{Interval $[0,\xi]$}
\label{sec-semicentre}
As $k(\nu,\xi)=0$ (see Eq.~(\ref{def-knu})) and $\lambda(\nu,\xi)=0$
at $\nu=\nu_0(\xi)$,  the solution of  Eq.~(\ref{eq-cauchyhalf}) has
thus the same form as for the infinite medium, namely
\begin{equation}  
j(\nu,\xi)=\xi\,p^\ast(\nu)\,P\frac{t(\nu,\xi)}{\lambda(\nu,\xi)} + 
f(\nu,\xi)\delta(\nu-\nu_0(\xi)),  
\label{eq-semicenter2}
\end{equation}
where $f(\nu,\xi)$ is still undetermined. An expression of
$f(\nu_0(\xi),\xi)$ is established when we consider the interval
$[\xi,\infty[$.

\subsubsection{Interval $[\xi,\infty[$}
\label{sec-semicut}
We now combine Eq.~(\ref{eq-semicenter2}) with
Eq.~(\ref{eq-cauchyhalf}). Replacing $\nu_0(\xi)$ by $\nu_0$ to simplify the
notation, we obtain
\begin{eqnarray}  
& &  \lambda(\nu,\xi)j(\nu,\xi)= \omega\,b\,\xi\,
  k(\nu,\xi)\pvint_{\xi}^\infty \frac{j(\nu',\xi)}{\nu'-\nu}\,d\nu'\nonumber\\
  & & +\  \omega\,b\,\xi\,k(\nu,\xi)\frac{f(\nu_0,\xi)}{\nu_0-\nu}
  + G(p^\ast,\nu,\xi).
\label{eq-semicauchy3}
\end{eqnarray}
In $G(p^\ast,\nu,\xi)$ we have regrouped the terms
depending on $p^\ast(\nu)$. After some simple algebra, which
makes use in particular of the definition of $\lambda(\nu,\xi)$ 
in Eq.~(\ref{def-lambdanuxi}), we can write
\begin{eqnarray}  
& & G(p^\ast,\nu,\xi)= {\xi}k(\nu,\xi)\pvint_{0}^{\xi}
\frac{p^\ast(\nu')}{\lambda(\nu',\xi)}\frac{d\nu'}{\nu'-\nu}\nonumber\\
& & + \ \xi\left[p^\ast(\nu)t(\nu,\xi) 
  + k(\nu,\xi)\pvint_{\xi}^\infty \frac{p^\ast(\nu')}{\nu'-\nu}\,d\nu'\right].
\label{def-semigbar1}  
\end{eqnarray}
The principal part in the integration over $[0,\xi]$ is required by
the zero of $\lambda(\nu,\xi)$ at $\nu=\nu_0(\xi)$.  In the
monochromatic limit, $G(p^\ast,\nu,\xi)$ depends only on the primary
source term $q^\ast(\nu)$, since $p^\ast(\nu)=q^\ast(\nu)$ when $b=1$
(see Eq.~(\ref{eq-defpstar1})).

To solve Eq.~(\ref{eq-semicauchy3}) for $j(\nu,\xi)$, we first
transform it into a boundary value problem in the complex plane.
We introduce the Hilbert transforms
\begin{equation}  
  N_{\rm c}(z,\xi)\equiv H_{\xi}^\infty[j(\nu,\xi)],\quad P_c(z,\xi)\equiv
  H_{\xi}^\infty[p^\ast(\nu)].   
\label{def-nxized2}
\end{equation}
The subscript c stands for continuous. In $P_c(z,\xi)$ the dependence
on $\xi$ comes from the integration domain. The two functions in
Eq.~(\ref{def-nxized2}) are analytic in the
    complex plane cut along $[\xi,\infty[$ and go to zero at infinity.

Introducing into Eq.~(\ref{eq-semicauchy3}) the Plemelj formulas for
${\cal L}(z,\xi)$, $N_c(z,\xi)$, and $P_c(z,\xi)$,
we obtain after some elementary algebra
\begin{equation}  
  {\cal L}^+(\nu,\xi) M^+(\nu,\xi) -  {\cal L}^-(\nu,\xi)M ^-(\nu,\xi)
  = \frac{1}{\omega\,b}m(\nu,\xi),
\label{eq-semicauchy6}  
\end{equation}
with
\begin{equation}  
m(\nu,\xi)= p^\ast(\nu)
-\  \left[{\cal L}^+(\nu,\xi) - {\cal
      L}^-(\nu,\xi)\right] \frac{1}{2\ui\pi}\pvint_{0}^{\xi}
\frac{p^\ast(\nu')}{\lambda(\nu',\xi)}\frac{d\nu'}{\nu'-\nu}.
\label{eq-semipmnuxi}
\end{equation}
Here $M^\pm(\nu,\xi)$ are the limiting values above and below the real
axis of the function $M(z,\xi)$ defined by
\begin{equation}  
 M(z,\xi)\equiv N_{\rm c}(z,\xi) +\frac{1}{2\ui\pi}
\frac{f(\nu_0,\xi)}{\nu_0-z} +\frac{1}{\omega\,b}P_{\rm c}(z,\xi).
\label{eq-defmnuxi}
\end{equation}
The functions  $M(z,\xi)$ and  ${\cal L}(z,\xi)$ have
different analyticity properties\,: $M(z,\xi)$ is analytic in the complex plane 
cut along $[\xi,\infty[$, while  the singular cut of ${\cal L}(z,\xi)$
is along $]-\infty,-\xi]\cup[+\xi,+\infty[$. A standard method for
solving such equations is to introduce an auxiliary function $X(z,\xi)$,  
with the same analyticity domain as the unknown function, here $M(z,\xi)$
\citep[see e.g.,][]{ablowitz97}.
This means that we have to construct a function $X(z,\xi)$
analytic in the complex plane cut along $[\xi,\infty[$.
In addition it should be free of singularities and solution to the
homogeneous Riemann--Hilbert problem 
\begin{equation} 
  \frac{X^+(\nu,\xi)}{X^-(\nu,\xi)}= \frac{{\cal
        L}^+(\nu,\xi)}{{\cal L}^-(\nu,\xi)}, \quad \nu\in[\xi,\infty[. 
\label{eq-rhmono}
\end{equation}
Taking the logarithm of this equation, making use of appropriate
Plemelj formulae and imposing the condition 
that $X(z,\xi)$ is free of singularity, one finds the exact expression
\begin{equation}  
  X(z,\xi)=\frac{1}{1-z}\exp\left[\frac{1}{\pi}\int_{\xi}^\infty\theta(\nu,\xi)
\frac{d\nu}{\nu-z}\right],
\label{eq-xzedmono}
\end{equation}
where $\theta(\nu,\xi)$ is the argument of ${\cal L}^+(\nu,\xi)$
\citep[e.g.,][p.~82]{musk53,gakhov66,ablowitz97,frisch22}. In the
following, we shall make use of the function
\begin{equation}  
  X^\ast(z,\xi)\equiv(\nu_0-z) X(z,\xi).
\label{eq-defxstar}
\end{equation}
It satisfies the factorization relation
\begin{equation}  
  X^\ast(z,\xi)X^\ast(-z,\xi)={\cal L}(z,\xi).
\label{eq-factorstar}
\end{equation}
We see here that the Wiener--Hopf factorization and the
Riemann--Hilbert problem stated in Eq.~(\ref{eq-rhmono}) are two sides of
the same coin. Combined with ${\cal L}(0,\xi)=\epsilon_b=1-\omega\,b$,
Eq.~(\ref{eq-factorstar}) leads to
$X^\ast(0,\xi)=\sqrt{\epsilon_b}$. In the monochromatic 
limit, i.e., when $b=\xi=1$, $X^\ast(z,\xi)$
is simply related to the monochromatic Chandrasekhar's 
H-function (see Section~\ref{sec-emergent}).  

To pursue the determination $M(z,\xi)$, we multiply
Eq.~(\ref{eq-semicauchy6}) by 
$(\nu_0-\nu)$. Making use of Eq.~(\ref{eq-rhmono}), we obtain
\begin{equation}  
[X^\ast]^+(\nu,\xi)M^+(\nu,\xi) - [X^\ast]^-(\nu,\xi)M^-(\nu,\xi)=
\frac{[X^\ast]^+(\nu,\xi)}
       {{\cal L}^+(\nu,\xi)}\,\frac{m(\nu,\xi)}{\omega\,b}, 
\label{eq-cauchy7}
\end{equation}
where $m(\nu,\xi)$ is defined in Eq.~(\ref{eq-semipmnuxi}) and
$M(z,\xi)$ in Eq.~(\ref{eq-defmnuxi}).

The left-hand side of this equation is the jump across
$[\xi,\infty[$ of the function $(\nu_0-z)X(z,\xi)M(z,\xi)$, analytic
in the complex plane cut along $[\xi,\infty[$.  The right-hand
side is the jump across $[\xi,\infty[$ of the function
\begin{equation}  
  {\cal M}(z,\xi)\equiv H_{\xi}^\infty\left[
\frac{[X^\ast]^+(\nu,\xi)}{{\cal L}^+(\nu,\xi)}\,m(\nu,\xi)\right].
\label{eq-defS1S2}
\end{equation}
The function ${\cal M}(z,\xi)$ has the standard properties of a Hilbert
transform, namely, it goes to zero at infinity and is free of singularities.  

We can now use the Liouville theorem to solve
Eq.~(\ref{eq-cauchy7}). We introduce   
\begin{equation}  
 F(z,\xi)\equiv(\nu_0-z)X(z,\xi)M(z,\xi)-\frac{1}{\omega\,b}{\cal M}(z,\xi).
\label{eq-semidefF0}
\end{equation}
By construction $F(z,\xi)$ is analytic in the complex plane cut along
the real semi-infinite line
$[\xi,\infty[$. According to Eq.~(\ref{eq-cauchy7}) its jump across 
$[\xi,\infty[$ is zero. It is also free of singularity. The factor
$(\nu_0-z)$ cancels the singularity in $M(z,\xi)$ due to the term
$f(\nu_0,\xi)/(\nu_0-z)$ (see Eq.~(\ref{eq-defmnuxi})). One may thus
        conclude that $F(z,\xi)$ is an entire 
function, which can be determined by its behavior at infinity.
Remembering that $X(z,\xi)\simeq -1/z$ as $z\to\infty$, we see that
all the terms in Eq.~(\ref{eq-semidefF0}) go to zero at infinity.
Hence
\begin{equation}  
  F(z,\xi)=0,
\label{eq-semidefF2}
\end{equation}
for all values of $z$ and $\xi$. We thus obtain
\begin{equation}  
   N_{\rm c}(z,\xi) = \frac{1}{\omega\,b}\frac{{\cal
       M}(z,\xi)}{(\nu_0-z)X(z,\xi)} 
   -\frac{1}{2\ui\pi}\frac{f(\nu_0,\xi)}{\nu_0-z}
   -\frac{1}{\omega\,b}P_{\rm c}(z,\xi).  
\label{eq-ncxisemi}
\end{equation}
Since $N_{\rm c}(z,\xi)$ has been defined as a Hilbert transform, it
should be free of singularity in its analyticity
domain. Equation~(\ref{eq-ncxisemi}) shows that $N_{\rm c}(z,\xi)$
has a singularity at $z=\nu_0$. It is readily eliminated by choosing
\begin{equation}  
  f(\nu_0,\xi)=\frac{2\ui\pi}{\omega\,b}\frac{{\cal
        M}(\nu_0,\xi)}{X(\nu_0,\xi)}. 
\label{eq-fnuzerosemi}
\end{equation}
The expressions of $N_{\rm c}(z,\xi)$ and $f(\nu_0,\xi)$ given above
still depend on $\bar\jmath(\nu)$.

For the infinite medium, we have seen in Section~\ref{sec-singular2}
that for the determination of $\bar\jmath(\nu)$ we need the integral of the
Hilbert transform of $j(\nu,\xi)$ over the full real axis. Here we
have to calculate $N(z,\xi)=H_0^{\infty}[j(\nu,\xi)]$,
the Hilbert transform of $j(\nu,\xi)$ over the full interval $[0,\infty[$. 
Summing the contributions from Eqs.~(\ref{eq-semicenter2}) and
(\ref{eq-ncxisemi}), we obtain 
\begin{eqnarray}  
& &  N(z,\xi)=\frac{1}{2\ui\pi}\int_0^\infty\frac{j(\nu,\xi)}{\nu
    -z}\,d\nu=
\frac{1}{\omega\,b}\frac{{\cal M}(z,\xi)}{X^\ast(z,\xi)} \nonumber\\
 & &\!\!\!\!\!\! -\frac{1}{2\ui\pi}\left\{\frac{1}{\omega\,b}\int_{\xi}^\infty
  p^\ast(\nu)\,\frac{d\nu}{\nu-z} - \xi\pvint_0^\xi
  p^\ast(\nu)\frac{t(\nu,\xi)}{\lambda(\nu,\xi)}\,\frac{d\nu}{\nu-z}\right\}. 
\label{eq-nzedtsemi2}
\end{eqnarray}
It is shown in Appendix~\ref{sec-jump} that 
\begin{eqnarray}  
& & {\cal M}(z,\xi)=
-X^\ast(z,\xi)\frac{1}{2\ui\pi}\pvint_{0}^{\xi} 
\frac{p^\ast(\nu)}{\lambda(\nu,\xi)}\,\frac{d\nu}{\nu-z}\nonumber\\
& & +\ \frac{1}{2\ui\pi}\int_{0}^{\infty}\frac{p^\ast(\nu)}
     {X^\ast(-\nu,\xi)}\,\frac{d\nu}{\nu-z}.   
\label{eq-s1s2}
\end{eqnarray}
Introducing ${\cal M}(z,\xi)$ into Eq.~(\ref{eq-nzedtsemi2}), using
$\lambda(\nu,\xi)=1-\omega\,b\xi\,t(\nu,\xi)$ and 
$p^\ast(\nu)=\omega(1-b)\bar\jmath(\nu) + q^\ast(\nu)$, we obtain the relation
\begin{equation}  
  \omega[ b\,N(z,\xi) + (1-b)\bar N(z)] + N_{\rm Q}(z)=\frac{1}{X^\ast(z,\xi)}
  \frac{1}{2\ui\pi}\int_0^\infty\frac{p^\ast(\nu)}
       {X^\ast(-\nu,\xi)}\,\frac{d\nu}{\nu-z}, 
\label{eq-semifull}
\end{equation}
where $\bar N(z)$ and $N_{\rm Q}(z)$ are the Hilbert transforms of
$\bar\jmath(\nu)$ and $q^\ast(\nu)$ defined by
\begin{equation}  
  \bar N(z)=H_0^\infty[\bar\jmath(\nu)],\quad \mbox{and}\quad
  \quad N_{\rm Q}(z)=H_0^\infty[q^\ast(\nu)]. 
\label{eq-semidefnbar}
\end{equation}
For the infinite medium, the equation corresponding to
Eq.~(\ref{eq-semifull}) is the much simpler
Eq.~(\ref{eq-relation}). We remark also that Eq.~(\ref{eq-semifull})
can be obtained without the separation in two intervals $[0,\xi]$ and
$[\xi,\infty[$, but we find that the need of a half-space auxiliary
    function appears more clearly when the intervals are treated
    separately. In the monochromatic limit $b=1$,
    Eq.~(\ref{eq-semifull}) provides an exact expression for
    $N(z,\xi)$ from which $j(\nu,\xi)$ can be derived by application
    of the Plemelj formulae. Equation~(\ref{eq-semifull}) becomes
also  simpler for the Green's function. When
    $Q^\ast(\tau)=\delta(\tau)$, the term $N_{\rm Q}(z)$ is replaced
    by $[[X^\ast(z,\xi)]^{-1} -1]/(2\ui\pi\omega)$ and $p^\ast(\nu)$
    by $\omega(1-b)\bar\jmath(\nu)$.

\subsection{The frequency integrated function ${\bar\jmath}(\nu)$}
\label{sec-semiintegral}
Proceeding as for the infinite medium, we integrate
Eq.~(\ref{eq-semifull}) over $\xi$. Making us of the factorization
relation in Eq.~(\ref{eq-factorstar}), we obtain
\begin{eqnarray}  
 & & \omega\bar N(z) + N_{\rm Q}(z)=\nonumber\\
  & & \frac{1}{2\ui\pi}\int_0^{\varphi_0}\frac{W(\xi)}{{\cal L}(z,\xi)}
\pvint_{0}^{\infty}\frac{X^\ast(-z,\xi)}{X^\ast(-\nu,\xi)}
  \,p^\ast(\nu)\,\frac{d\nu}{\nu-z}\,d\xi.   
\label{eq-semijbar1}
\end{eqnarray}
The singular integral in the right-hand side of Eq.~(\ref{eq-semijbar1})
cannot be expressed in terms of the Hilbert transform of the unknown
function $p^\ast(\nu)$ because of the factor $1/X^\ast(-\nu,\xi)$. 
A method for solving singular integral equations with Cauchy-type
kernels is described in \citet[Chapter 6]{musk53}. Following this
method, also applied in \citet{hemsch72},
the kernel is separated
into a ``dominant'' part, containing the Cauchy singularity $1/(\nu-z)$,
and a ``Fredholm'' part. This amounts to
introduce a function $r(z,\nu,\xi)$ defined by
\begin{equation}  
\frac{X^\ast(-z,\xi)}{X^\ast(-\nu,\xi)}\,
\frac{1}{\nu-z} =\frac{1}{\nu-z} + r(z,\nu,\xi).
\label{eq-kernelgene}
\end{equation}
Regrouping the integrals with Cauchy kernels and writing now $\Omega(z)$ as
\begin{equation}  
  \Omega(z)\equiv 1-\frac{1-b}{b}\int_0^{\varphi_0}W(\xi)\left[
  \frac{1}{{\cal L}(z,\xi)}-1\right]\,d\xi,
\label{def-omega3}
\end{equation}
we can rewrite Eq.~(\ref{eq-semijbar1})  as
\begin{equation}  
\Omega(z)\left[\bar N(z) + \frac{1}{\omega(1-b)}N_{\rm Q}(z)\right]=G(z) + R(z),
\label{eq-semijbar2} 
\end{equation}
where
\begin{equation}  
   R(z)\equiv\frac{1-b}{b}\int_0^\infty{\bar\jmath}(\nu)V(z,\nu)\,d\nu,
\label{eq-defgrdm}
\end{equation}
\begin{equation}  
G(z)\equiv \frac{1}{\omega(1-b)}N_{\rm Q}(z)
+ \frac{1}{\omega\,b}\int_0^\infty q^\ast(\nu)V(z,\nu)\,d\nu,
\label{eq-defggdm}
\end{equation}
and
\begin{equation}  
  V(z,\nu)\equiv\frac{1}{2\ui\pi}\int_0^{\varphi_0}\frac{W(\xi)}{{\cal
    L}(z,\xi)}r(z,\nu,\xi)\,d\xi.
\label{eq-defvznu}
\end{equation}
The term $R(z)$, which involves $\bar j(\nu)$ and the Fredholm part of
the kernel, will prevent us from constructing with Eq.~(\ref{eq-semijbar2})
an exact expression for $\bar\jmath(\nu)$. It will however let us
construct for ${\bar\jmath}(\nu)$ a Fredholm integral equation. 

\subsection{A Fredholm integral equation for ${\bar\jmath}(\nu)$}
\label{sec-fredholm}
In Eq.~(\ref{eq-semijbar2}) $\bar N(z)$ is analytic in the complex
plane cut along $[0,\infty[$, while $\Omega(z)$, $R(z)$
and $G(z)$ are analytic in $\cset\ /\rset$.
Thus $\bar N(z)$ cannot be derived directly from this
equation. However, it can provide for $\nu\in[0,\infty[$ a
boundary value problem, namely
\begin{eqnarray}  
& & \Omega ^+(\nu)[\bar N^+(\nu) + \frac{1}{\omega(1-b)}N_{\rm
      Q}^+(\nu)]\nonumber\\
& &  -\ \Omega ^-(\nu)[\bar N^-(\nu) +
    \frac{1}{\omega(1-b)}N_{\rm Q}^-(\nu)] =\rho(\nu) + g(\nu),
\label{eq-semicut}
\end{eqnarray}
with 
\begin{equation}  
 \rho(\nu)\equiv R^+(\nu) - R^-(\nu),\quad
 g(\nu)\equiv G^+(\nu) - G^-(\nu).  
\label{eq-defrhogstar}
\end{equation}
Equation~(\ref{eq-defgrdm}) shows that $\rho(\nu)$ may be written as
\begin{equation}  
  \rho(\nu)=\frac{1-b}{b}\pvint_0^\infty
      {\bar\jmath}(\nu'){\cal V}(\nu,\nu')\,d\nu', 
\label{eq-finalrhot}
\end{equation}
where ${\cal V}(\nu,\nu')$ is the jump of the function $V(z,\nu)$ defined in
Eq.~(\ref{eq-defvznu}). This jump is calculated in Appendix~\ref{sec-jump}
(see Eqs.~(\ref{eq-defcalv1}), (\ref{eq-defcalv2}), and
(\ref{eq-defgrv})). Equation~(\ref{eq-defggdm}) shows 
that the jump $g(\nu)$ may be written as 
\begin{equation}   
  g(\nu)= \frac{1}{\omega(1-b)}q^\ast(\nu)
  +\frac{1}{\omega\,b}\pvint_0^\infty 
      q^\ast(\nu'){\cal V}(\nu,\nu')\,d\nu'.
\label{eq-defgstar}
\end{equation}
For complete frequency redistribution $r(z,\nu,\xi)$ is zero, hence
$\rho(\nu)=0$ and $g(\nu)$ is reduced to its first term.

The problem is now to solve Eq.~(\ref{eq-semicut}) for $\bar
N(z)$. Because the analyticity domain of $\Omega(z)$ is cut along the
real axis, the solution of this equation requires the introduction
of an auxiliary function, denoted $\bar X(z)$, analytic in the complex
plane cut along $[0,\infty[$, free of zero and singularity, and such that
\begin{equation}  
 \frac{\bar X^+(\nu)}{\bar X^-(\nu)}=\frac{\Omega^+(\nu)}{\Omega^-(\nu)},\quad
 \nu\in[0,\infty[.
\label{eq-defygrec}
\end{equation}
As shown in Section~\ref{sec-omega1}, $\Omega(z)$ is free of
zeroes. Hence $\bar X (z)$ is given by
\begin{equation}  
   \bar X(z)\equiv\exp\left[\frac{1}{\pi}\int_{0}^\infty{\bar\theta}(\nu)
\frac{d\nu}{\nu-z}\right], 
\label{eq-yzedcr}
\end{equation}
where ${\bar\theta}(\nu)$ is the argument of $\Omega ^+(\nu)$. The
definition of $\bar X(z)$ shows that $\bar X(z)\to 1$ as
$z\to\infty$. The function $\bar X(z)$ satisfies the factorization relation
\begin{equation}  
  \bar X(-z)\bar X(z)=\Omega(z),
\label{eq-factorxcr}
\end{equation}
for $z$ outside the cut $[0,\infty[$. In the limit $b=0$, $\Omega(z)$
becomes identical to the dispersion function for complete frequency
redistribution ${\cal L}_{\rm R}(z)$ (see Section~\ref{sec-omega2})
and the function $\bar X(z)$ identical to the $X$-function for
complete frequency redistribution, henceforth denoted $X_{\rm R}(z)$.

Knowing that $\Omega(z)$ is free of zeroes, we can divide
Eq.~(\ref{eq-semicut}) by $\Omega ^+(\nu)$ and make use of
Eq.~(\ref{eq-defygrec}). We also introduce the two Hilbert transforms
\begin{equation}  
  {\cal G}(z)\equiv H_0^\infty\left[\frac{\bar X^+(\nu)}{\Omega
    ^+(\nu)}g(\nu)\right] =
  H_0^\infty\left[\frac{g(\nu)}{\bar X(-\nu)}\right], 
\label{eq-defcalg}
\end{equation}
\begin{equation}
  {\cal R}(z)\equiv H_0^\infty\left[\frac{\bar X^+(\nu)}{\Omega
    ^+(\nu)}\rho(\nu)\right] =  H_0^\infty\left[\frac{\rho(\nu)}{\bar
      X(-\nu)}\right]. 
\label{eq-defcalm}
\end{equation}
They are analytic in the complex plane cut along
$[0,\infty[$. We can now write Eq.~(\ref{eq-semicut}) as 
\begin{eqnarray}  
& & \!\!\!\!\! \bar X^+(\nu)\left[\bar N^+(\nu) + \frac{N^+_{\rm
        Q}(\nu)}{\omega(1-b)}\right] - 
  \bar X^-(\nu)\left[\bar N^-(\nu) + \frac{N^-_{\rm
        Q}(\nu)}{\omega(1-b)}\right]\nonumber\\ 
& & =\ [{\cal G}^+(\nu) - {\cal G} ^-(\nu)]+ [{\cal R}^+(\nu)-{\cal
      R}^-(\nu)].   
\label{eq-semicuthalf}
\end{eqnarray}
Following the method already used several times, we introduce
\begin{equation}  
  \bar F(z)\equiv \bar X(z)\left[\bar N(z)
    +\frac{N_{\rm Q}(z)}{\omega(1-b)}\right] -{\cal G}(z)-{\cal R}(z). 
\label{eq-semidefFbar}
\end{equation}
By construction $\bar F(z)$ is analytic in the complex plane cut along
$[0,\infty[$ and according to Eq.~(\ref{eq-semicuthalf}) its jump
    along this cut is zero. Hence it is an entire function. It is also
    easy to see that it goes to zero at infinity. We can thus conclude
    that ${\bar F}(z)$ is zero for all $z$. We thus obtain
\begin{equation}  
  \bar N(z)  + \frac{1}{\omega(1-b)}N_{\rm Q}(z)=\frac{1}{\bar
    X(z)}\left[{\cal G}(z)+{\cal R}(z)\right]. 
\label{eq-seminbarzedg}
\end{equation}    
Applying the Plemelj formulae to all the terms in this equation and
using the expression of $\rho(\nu)$ given in Eq.~(\ref{eq-finalrhot}), we
obtain for $\bar\jmath(\nu)$ a Fredholm integral equation, which may
be written as 
\begin{equation}  
  \bar\jmath(\nu) -
  \int_0^\infty\bar\jmath(\nu')\kappa(\nu,\nu')\,d\nu'=
  -\frac{1}{\omega(1-b)}q^*(\nu) + \eta(\nu). 
\label{eq-fredholmhf}
\end{equation} 
The kernel $\kappa(\nu,\nu')$ is given by 
\begin{eqnarray}  
& & \kappa(\nu,\nu')= \frac{1}{2}\left[\frac{1}{\Omega^+(\nu)} +
    \frac{1}{\Omega^-(\nu)}\right] {\cal V}(\nu,\nu')\nonumber\\
& & + \left[\frac{1}{\bar X^+(\nu)} -\frac{1}{\bar X^-(\nu)}\right]
 \frac{1}{2\ui\pi}\pvint_0^\infty\frac{{\cal V}(\nu'',\nu')}{\bar X(-\nu'')}
\,\frac{d\nu''}{\nu''-\nu},
\label{eq-defkappa}
\end{eqnarray}
and $\eta(\nu)=[{\cal G}(z)/\bar X(z)]_{\rm jmp}$ by
\begin{eqnarray}  
& & \eta (\nu)=\frac{1}{2}\left[\frac{1}{\Omega^+(\nu)} +
    \frac{1}{\Omega^-(\nu)}\right]\,g(\nu)\nonumber\\
& & + \left[\frac{1}{\bar X^+(\nu)} -\frac{1}{\bar X^-(\nu)}\right]
 \frac{1}{2\ui\pi}\pvint_0^\infty\frac{g(\nu')}{\bar
   X(-\nu')}\,\frac{d\nu'}{\nu'-\nu}.   
\label{eq-inhomot}
\end{eqnarray}
An explicit expression of ${\cal V}(\nu,\nu')$ is given in
Appendix~\ref{sec-jump} and $g(\nu)$ is defined in
Eq.~(\ref{eq-defgstar}). Using the definition of $\bar X^\pm(\nu)$ in 
Eq.~(\ref{eq-defygrec}) and the factorization relation in
Eq.~(\ref{eq-factorxcr}), it is possible to express all the square
brackets in terms of $\bar X(-\nu)$ and of the imaginary and real
parts of $\Omega ^+(\nu)$.

For the Green's function, that is for $Q^\ast(\tau)=\delta(\tau)$,
Eq.~(\ref{eq-fredholmhf}) becomes
\begin{equation}  
  \bar\jmath(\nu) -
  \int_0^\infty\bar\jmath(\nu')\kappa(\nu,\nu')\,d\nu'=\eta_{\rm g}(\nu), 
\label{eq-fredholmg}
\end{equation}
with
\begin{equation}  
  \eta_{\rm g}(\nu)=\frac{1}{2\ui\pi(1-b)}\left[\frac{1}{\bar X^+(\nu)}-
    \frac{1}{\bar X^-(\nu)}\right] + \left[\frac{{\cal G}_{\rm
          g}(z)}{\bar X(z)}\right]_{\rm jmp}.
\label{eq-defetag}
\end{equation}
The function ${\cal G}_{\rm g}(z)$ is a Hilbert transform defined as
in Eq.~(\ref{eq-defcalg}), the function $g(\nu)$  being replaced by
\begin{equation}  
  g_{\rm g}(\nu)\equiv\frac{1}{2\ui\pi\,b}\!\!\int_0^{\varphi_0}\!
\left[\frac{W(\xi)}{{\cal L}(z,\xi)}\right]_{\rm
  jmp}[(\nu_0(\xi)+\nu)X(-\nu,\xi)-1]\,d\xi.
\label{eq-defgnuapp}
\end{equation}

The equation for ${\bar j}(\nu)$ becomes of course much simpler for
complete frequency redistribution and provides an explicit exact
solution. First, we have $\kappa(\nu,\nu')=0$. As for $\eta(\nu)$ it is
given by Eq.~(\ref{eq-inhomot}) in which $g(\nu)$ is replaced by
$q^\ast(\nu)/(\omega(1-b))$. A further
simplification appears when we consider the Green's function. Indeed,
when $b=0$, we have  ${\cal L}(z,\xi)=1$. Hence the jump of ${\cal
  L}(z,\xi)$ is zero, $g_{\rm g}(\nu)$ is also zero and $\eta_{\rm
  g}(\nu)$ is reduced to its first term. We thus obtain
\begin{equation}  
  \bar\jmath_{\rm R}(\nu)=\frac{1}{\omega\,2\ui\pi}\left[\frac{1}{
      X^+_{\rm R}(\nu)}-
    \frac{1}{X^-_{\rm R}(\nu)}\right]=
\frac{k_{\rm R}(\nu)}{{\cal L}^+_{\rm R}(\nu)
{\cal L}^-_{\rm R}(\nu)}X_{\rm R}(-\nu).
\label{eq-resolvent}
\end{equation} 
The Laplace transform of $\bar\jmath_{\rm R}(\nu)$ yields the
resolvent function for complete frequency redistribution in a
semi-infinite medium \citep[see e.g.,][p.~104]{ivanov73,frisch22}.

\subsection{The emergent intensity}
\label{sec-emergent}
The emergent intensity for a uniform primary source is studied in
detail, analytically and numerically in \citet{hemsch72}, for a
frequency dependent branching ratio $b$.
This article shows in particular how the line profile is affected by
a monochromatic component acting at large frequencies. Their results
are discussed in some detail in Section~\ref{sec-conclusion}. Here we
present some exact expressions for an arbitrary primary source
$Q^\ast(\tau)$ and a constant branching ratio $b$.

\begin{figure}
\begin{center}
\includegraphics[scale=0.55]{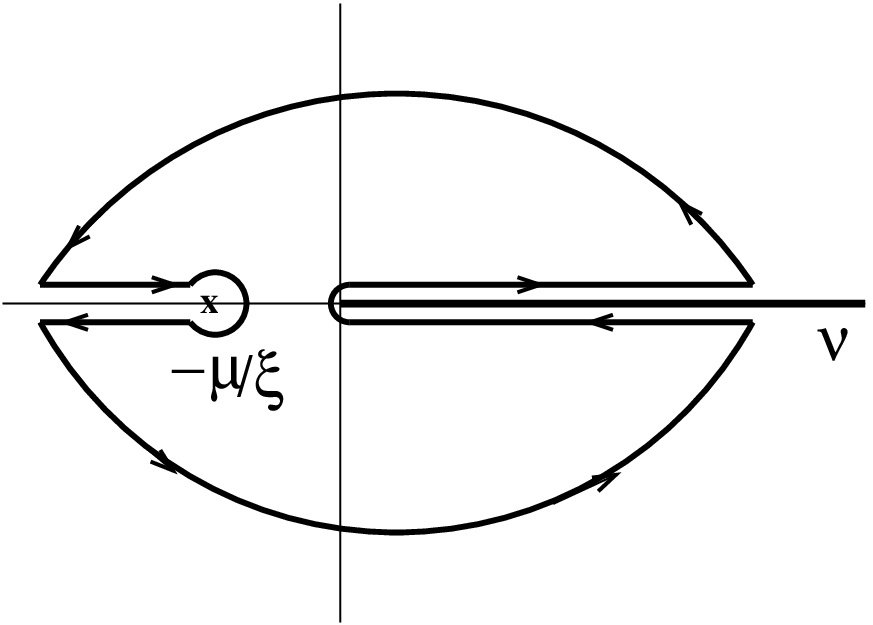}
\end{center}
\caption{Contour in the complex plane for the calculation of the
  emergent intensity. The function $N_{\rm S}(z,\xi)$ is analytic in the
  complex plane cut along the semi-infinite real line $[0,\infty[$.}
\label{fig-intensity-contour}
\end{figure}

In terms of the source function $S(\tau,\xi)$, the emergent intensity
is given by 
\begin{equation}  
  I(0,\xi,\mu)=\int_0^\infty S(\tau,\xi)\,\ue
  ^{-\tau\xi/\mu}\frac{\xi}{\mu}\,d\tau. 
\label{eq-iemergent}
\end{equation}
We recall that $\xi=\varphi(x)$ and that $\mu$ is the cosine
of the radiation field direction. The source function $S(\tau,\xi)$, 
defined in Eq.~(\ref{def-source}), may be written as
\begin{equation}  
  S(\tau,\xi)=\int_0^\infty s(\nu,\xi)\,\ue ^{-\nu\tau}\,d\nu,
\label{eq-sourcehalf}
\end{equation}
where $s(\nu,\xi)$ is given by
\begin{equation}
  s(\nu,\xi)=\omega[b\,j(\nu,\xi) + (1-b){\bar\jmath}(\nu)]
  + q^\ast(\nu).
\label{eq-defsnug}
\end{equation}
Combining Eqs.~(\ref{eq-iemergent}) and (\ref{eq-sourcehalf}) and
integrating over $\tau$, we obtain
\begin{equation}  
  I(0,\xi,\mu)=\int_0^\infty\frac{\xi}{\xi + \mu\nu} s(\nu,\xi)\,d\nu.
\label{eq-iemergent2}
\end{equation}
To calculate this integral we introduce $N_{\rm S}(z,\xi)$ the Hilbert
transform of $s(\nu,\xi)$, denoted $N_{\rm S}(z,\xi)$. According
to Eq.~(\ref{eq-defsnug}), we have
\begin{equation}  
N_{\rm S}(z,\xi)\equiv\omega\left[b N(z,\xi) + (1-b)\bar N(z)\right] +
N_{\rm Q}(z), 
\label{eq-nszedxi}
\end{equation}
where $N(z,\xi)$, $\bar N(z)$ and $N_{\rm Q}(z)$ are the Hilbert
transform over $[0,\infty[$ of $j(\nu,\xi)$, $\bar\jmath(\nu)$, and
$q^\ast(\nu)$. These functions are  
analytic in the complex plane cut along the positive real axis,
free of singularities, and go to zero at infinity. The same properties
hold true for $N_{\rm S}(z,\xi)$. Applying the Plemlej formulae to $N_{\rm
  S}(z,\xi)$, we can write the emergent intensity as
\begin{equation}  
  I(0,\xi,\mu)=\int_0^\infty\frac{\xi}{\xi + \mu\nu} [N_{\rm S}^+(\nu,\xi)
  - N_{\rm S}^-(\nu,\xi)]\,d\nu,
\label{eq-iemergentcr}
\end{equation}
where $N_{\rm S}^\pm(\nu,\xi)$ are the limiting values of $N(z,\xi)$
above and below the positive real axis.  Integrating over the contour shown in
Figure~\ref{fig-intensity-contour}, which circulates around the cut of $N_{\rm
  S}(z,\xi)$ and around the pole at $z=-\xi/\mu$, and remembering that
$N_{\rm S}(z,\xi)$ goes to zero at infinity, we find
\begin{equation}  
  I(0,\xi,\mu)=2\ui\pi\frac{\xi}{\mu}N_{\rm S}(-\frac{\xi}{\mu},\xi).
\label{eq-iemergentgene}
\end{equation}
The expression of $N(z,\xi)$ given in Eq.~(\ref{eq-semifull}) leads to
\begin{equation}  
N_{\rm S}(z,\xi)=
\frac{1}{X^\ast(z,\xi)}
\frac{1}{2\ui\pi}\int_0^\infty\!\!\frac{\omega(1-b)\bar\jmath(\nu)
+ q^\ast(\nu)}{X^\ast(-\nu,\xi)}\,\frac{d\nu}{\nu-z},
\label{eq-sourcehilbert}
\end{equation}
where $\bar\jmath(\nu)$ is the solution of the Fredholm integral
equation in Eq.~(\ref{eq-fredholmhf}). There is thus no explicit
expression for $N_{\rm S}(z,\xi)$ nor for the emergent intensity.
In the limiting cases of monochromatic scattering and complete
frequency redistribution, Eq.~(\ref{eq-sourcehilbert}) provides
well known results.

For {\bf monochromatic scattering}, $b=1$, and  $\xi$ has a constant
value, which we take equal to one, following a usual practice. 
Equation~(\ref{eq-sourcehilbert}) straightforwardly leads to 
\begin{equation}  
 N_{\rm S}(z,\xi)=\frac{1}{{X}^\ast_{\rm M}(z)}\frac{1}{2\ui\pi}
\int_0^\infty\frac{1}{
  X^\ast_{\rm M}(-\nu)}q^\ast(\nu)\,\frac{d\nu}{\nu-z},   
\label{eq-defnsmonog}
\end{equation}
where $X^\ast_{\rm M}(z)=X^\ast(z,1)=(\nu_0(1)-z)X(z,1)$.

For {\bf complete frequency redistribution},
$b=0$. Equation~(\ref{eq-nszedxi}) leads to 
\begin{equation}  
  N_{\rm S}(z,\xi)=\omega\bar N(z) + N_{\rm Q}(z).
\label{eq-nsource0}
\end{equation}
The expression of $\bar N(z)$ for complete frequency redistribution 
can be derived from Eq.~(\ref{eq-seminbarzedg}). Setting ${\cal
  R}(z)=0$ and $g(\nu)=q^\ast(\nu)/(\omega(1-b))$ in the
definition of ${\cal G}(z)$ (see Eq.~(\ref{eq-defcalg})), we readily obtain
\begin{equation}
N_{\rm S}(z,\xi)=\frac{1}{X_{\rm R}(z)}\frac{1}{2\ui\pi}
\int_0^\infty\frac{1}{
  X_{\rm R}(-\nu)}q^\ast(\nu)\,\frac{d\nu}{\nu-z}.
\label{eq-nsourcerc}
\end{equation}
We have the same expression as for monochromatic scattering,
with $X^\ast_{\rm M}(z)$, replaced by $X_{\rm R}(z)$. For monochromatic
scattering and complete frequency redistribution, $N_{\rm S}(z,\xi)$
depends only on $z$, a result which is consistent with the fact that the
source function depends only on $\tau$.
In these two limits, the emergent intensity may thus be written as
\begin{equation}  
I(0,\xi,\mu)=H(\frac{\mu}{\xi})\int_0^\infty \frac{\xi}{\xi 
    +\mu\nu} q^\ast(\nu)H(\frac{1}{\nu})\,d\nu, \quad \mu\ge 0,
\label{eq-iemergentcr23}
\end{equation}
with $\xi=\varphi(x)$ for complete frequency redistribution  and
$\xi=1$ for monochromatic scattering. The $H$-function is defined by
\begin{equation}  
 H(z)=\left\{\begin{array}{c}
1/X_{\rm R}(-1/z), \quad \mbox{complete redistribution},\\
1/X_{\rm M}^\ast(-1/z), \quad \mbox{monochromatic scattering.}
\end{array}\right.
\label{eq-defh}
\end{equation}
For monochromatic scattering, $H(z)$ is
the Chandrasekhar $H$-function for isotropic scattering \citep{chandra60}.
For complete frequency redistribution,
the properties of the $H$-function are discussed in detail in, e.g.,
\citet{ivanov73} \citep[see also][]{frisch22}. They are similar to those of the
monochromatic $H$-function. For example, they both increase
monotonically from $H(0)=1$ to $1/\sqrt{\epsilon}$ at infinity. The
value at infinity is easily derived from the factorization relations
for $X_{\rm R}(z)$ and $X^\ast_{\rm M}(z)$ and from ${\cal L}_{\rm
  M}(0)={\cal L}_{\rm R}(0)=\epsilon$. 

Equation~(\ref{eq-iemergentcr23}) is valid for all positive values of
$\mu$, but only those belonging to the interval $[0,1]$ are needed for
the calculation of the emergent intensity. This equation provides a very
convenient expression for the emergent intensity, especially when
$q^\ast(\nu)$ has a simple form. For example, when the primary source
has a uniform value $Q^\ast$, its inverse Laplace transform is
$q^\ast(\nu)=Q^\ast\delta(\nu)$. In the monochromatic and complete
frequency redistribution limits, the emergent intensity is thus simply
given by
\begin{equation}  
  I(0,x,\mu)=\frac{Q^\ast}{\sqrt{\epsilon}}H(\frac{\mu}
{\varphi(x)}),   
\label{eq-iemergentunif}
\end{equation}
with $\varphi(x)=1$ in the monochromatic case. 

Everywhere in this article it has been assumed implicitly that
$\epsilon\ne 0$. The results that have been obtained 
 hold also for $\epsilon=0$. Indeed, provided $b\ne 1$,
the parameter $\epsilon_b$ will remain strictly positive, smaller
than one and the dispersion function ${\cal L}(z,\xi)$ will have two
zeroes in the interval $[-\xi, +\xi]$. Only in the monochromatic limit
$b=1$ will there be a difference, since ${\cal L}(z,\xi)$ will have a double
zero at the origin.

\section{Summary of the solution method and concluding remarks}
\label{sec-conclusion}

Although the partial frequency redistribution model considered here is
fairly simple, namely a linear combination of monochromatic scattering
and complete frequency redistribution, the task of constructing exact
solutions is not straightforward, due to the coupling between the two
different scattering mechanisms. Two successive steps are needed,
  whatever the geometry. The first one requires the solution of a
monochromatic scattering problem and the second one of a complete
frequency redistribution one. For an infinite medium, the two steps
can be solved explicitly, making it possible to construct an exact
solution for the full partial frequency redistribution problem. For a
semi-infinite medium, the second step has no explicit solution but can
be recast as a Fredholm integral equation, which will require a
numerical solution. These results, first established in
\citet{hemsch71} and \citet{hemsch72} are confirmed in this article.
In \citet{hemsch72}, the singular eigenfunction expansion method
introduced by \citet{case60} is applied to the homogeneous radiative
transfer equation. The boundary conditions and known inhomogeneous
terms, coming from primary source of photons, are then taken into
account to construct the final solution. The method described here
does not require the introduction of singular eigenfunctions and
directly provides the final solution. The radiative transfer equation
is first transformed into an convolution integral equation, which
takes into account primary source terms and boundary conditions. The
convolution equation is then transformed into a singular integral
equation with an inverse Laplace transform, more or less equivalent to
the singular eigenfunction expansion. Both methods lead to a
  singular integral equation, one of them taking into account internal
  primary source terms, and boundary conditions, the other not. They
  are both solved by transforming them into a boundary value problem
  in the complex plane.

The solution of the partial frequency redistribution problem depends
critically on the branching ratio $b$ between complete frequency
redistribution for $b=0$ and monochromatic scattering for $b=1$.
Monochromatic scattering and complete frequency
redistribution have well understood but quite different large scale
asymptotic properties. 
For monochromatic scattering, the kernel $K_{\rm M}(\tau)$ has a
finite second order moment and the dispersion function has isolated
zeroes. This implies that the large scale behavior of the radiative
transfer equation is a diffusion equation and also that the
monochromatic Green's function in an infinite medium decreases at
infinity as $\exp(-\nu_0|\tau|)$, where $\nu_0$ is the positive root
of the dispersion function ${\cal L}_{\rm M}(z)$. In contrast, for
complete frequency redistribution, the kernel $K_{\rm R}(\tau)$
defined in Eq.~(\ref{eq-defkernelrc}) has an infinite second order
moment, the dispersion function ${\cal L}_{\rm R}(z)$ has no zero, and
the large scale behavior of the radiation field has the structure of a
L\'evy walk. More specifically the large scale transport equation
contains a Laplacian with a fractional power, equal to 1/2 or 1/4
depending on whether $\varphi(x)$ is a Doppler or a Voigt profile. The
infinite medium Green's function decreases then algebraically at
infinity as $1/\tau ^2$ for a Doppler profile and as $1/\tau ^{3/2}$
for a Voigt profile\citep[see e.g.,][]{ivanov73,frisch22}.

Whether the transition between the complete frequency redistribution
and the monochromatic regime is rapid or slow and for which values of
the branching ratio $b$ it occurs is an interesting question. The
exact expression  for $S(\tau,\xi)$ given in Eq.~(\ref{eq-sourcegene})
suggests that the 
overall asymptotic behavior at large $\tau$ will be dominated by
complete frequency redistribution contributions, unless $b$ is close
to one. Indeed, as soon as $b\ne 1$, Eq.~(\ref{eq-sourcegene}) shows that
there will be contributions from $\bar\jmath(\nu)$ and $j_{\rm
  d}^{(1)}(\nu,\xi)$ for $\nu\in[0,\xi]$. That the
large $\tau$ behavior is dominated by complete frequency
redistribution when there is a combination of monochromatic scattering
and complete frequency redistribution is also suggested by
a large $\tau$  asymptotic analysis of the source
function $S(\tau,\xi)$ \citep[see e.g.,][p.~569]{frisch22}.   
This conjecture is also in agreement with the numerical investigation of 
the emergent intensity reported in \citet{hemsch71} and \citet{hemsch72}.
Line profiles of the emergent intensity for a semi-infinite medium
with a uniform primary source are calculated for a frequency dependent
parameter $b$ equal to 0 in the line center and equal to some constant
value $b_{\rm W}$ in the wings.  Numerical results for $b_{\rm
  W}=0.2,0.5,0.9$ show that $b_{\rm W}$ has to be close to 1 for
departures from the complete frequency redistribution limit to become
significant. Having been obtained with a model which does preserve
energy conservation, these results may not be fully trustworthy.

A numerical solution of this partial frequency redistribution problem
for different values of the parameter $b$ appears to be the best way
to analyze the transition between the two asymptotic regimes. Although
\citet{hemsch71} has been able to use the exact 
expressions for the calculation of line profiles, a direct numerical
solution of the radiative transfer equation appears easier.  Some
precautions should however be taken. To be able to observe an
asymptotic regime, the number of scatterings undergone by the photons
should be large enough, which means $\epsilon$ small enough, say of
order $10^{-4}$. The total optical depth of the medium, say $\tau_{\rm
  max}$, should be roughly ten times the thermalization length
$\tau_{\rm th}$. This length scales as $1/\sqrt{\epsilon}$ for
monochromatic scattering. For complete frequency redistribution it
scales as $1/\epsilon$ (neglecting a logarithmic factor) for a Doppler
profile and as $a/\epsilon^2$ for a Voigt profile. Here $a$, the Voigt
parameter of the line, has typical values around $10^{-3}$. For
complete frequency redistribution, the frequency range should be wide
enough to ensure that the last frequency point, say $x_{\rm max}$,
satisfies $\varphi(x_{\rm max})\tau_{\rm th}\ll 1$, otherwise there is
a spurious accumulation of photons \citep{avrett65}. With these
constraints is seems better to avoid Monte--Carlo simulations. There
is no lack of standard numerical schemes to solve partial frequency
redistribution problem. A possibility is an Accelerated
$\Lambda$-Iteration (ALI) method developed by \citet{paletou95}. 

\section{Acknowledgements}
The author is very grateful to Dr. M. Hemsch for having introduced her
to this radiative transfer problem with partial frequency
redistribution and for having provided a copy of his PhD thesis. She
is also very thankful for many stimulating discussions. The author is
also very grateful to the referees for asking several clarification questions
and in particular to Professor Gary Ganapol for showing how to solve
the infinite medium problem with a Fourier transform applied directly to
the radiative transfer equation.

\begin{appendix}
\numberwithin{equation}{section}

\section{Remarks on the partial frequency redistribution model}
\label{sec-kneer}

A linear combination of complete frequency
redistribution and monochromatic scattering was introduced in
\citet{jefferies60} to model the formation of spectral lines, known as
resonance lines. These lines are formed by radiative transitions between
an infinitely thin fundamental energy level and a naturally broadened
excited one. In the rest frame of the atom, the scattering is coherent 
in frequency and the absorption profile is a Lorentzian. In
the laboratory frame, when the Doppler broadening is taken into
account, the absorption profile becomes a Voigt function and partial
frequency redistribution sets in. Explicit expressions for this
partial frequency redistribution mechanism known as R$_{\rm II}$ have
been established by \citet{hummer62} \citep[see
  also][]{hubeny15,frisch22}. As can be observed, for example in
Figure~1 of \citet{jefferies60}, R$_{\rm II}$ has a complete frequency
redistribution behavior in the line core and a monochromatic one in
the line wings.  In \citet{jefferies60}, this behavior was estimated
too complex for application and it was suggested to have $b=0$
in the line core, $b=1$ in the lines wings, with the jump from 0 to 1
being positioned at a few Doppler widths. It is shown in
\citet{kneer75} that this approximation does not meet essential properties
of a scattering process. Indeed a redistribution function of the form
\begin{equation}  
  r(x,x')=b(x)\varphi(x)\delta(x-x')  + [1-b(x)]\varphi(x)\varphi(x'),
\label{eq-redistr2}
\end{equation}
as used in \citet{hemsch72} does not satisfy the conservation of the
photons. In other terms, when $\epsilon=0$ and $Q^\ast(\tau)=0$ the
radiative flux should be a constant, since the medium is conservative
and free of primary sources. Integrating the left-hand side of
the radiative transfer equation over directions and frequencies, we
see that the derivative of the radiative flux is zero, if and only if
$b$ is a constant. Moreover $r(x,x')$ is not symmetric in $x$ and $x'$,
unless $b$ is a constant.  A simple cure to the problem is proposed in
\citet{kneer75}, but it requires that the parameter $b$ be a function
of $x$ and $x'$.  Nowadays the explicit expressions given in
\citet{hummer62} for the R$_{\rm II}$ redistribution function are
commonly used in numerical work and efficient numerical methods have
been developed to handle partial frequency redistribution \citep[see
  e.g.][]{hummer69,hubeny15}. Comparisons with the predictions of the
Kneer's approximation show that it fails in the line wings at
reproducing the intensity and the linear polarization created by the
scattering process. The reason is that it ignores the Doppler
broadening of the monochromatic component \citep{faurob88}.


\section{From a convolution integral equation to a singular integral
  equation}
\label{sec-whtocauchy}

We show here how to transform the convolution equation for ${\cal
  J}(\nu,\xi)$ given in Eq.~(\ref{eq-avintbis2}) into a singular integral
equation for $j(\nu,\xi)$, its inverse Laplace transform defined in
Eq.~(\ref{eq-inverse1}).  The method 
described below was proposed for complete frequency redistribution in
\citet{frifri82} and for monochromatic scattering in
\citet{frisch88} \citep[see also][]{frisch22}.  Its goal is to write
each of the integrals over $\tau'$ in the right-hand side of
Eq.~(\ref{eq-avintbis2} as a Laplace transform. To fix the idea we 
assume $\tau>0$. Let us consider the integral
\begin{equation}  
  \int_{-\infty}^{+\infty}K_{\rm M}(|\tau-\tau'|\xi){\cal J}(\tau',\xi)\,d\tau'.
\label{eq-rhs}
\end{equation}
The integration over $\tau'$ is divided into three integrals\,: 
\begin{equation}  
 \int_{-\infty}^{+\infty}d\tau'=\int_{-\infty}^0\,d\tau' +
 \int_0^\tau \,d\tau' + \int_{\tau}^{+\infty} \,d\tau'= [1] +[2] +[3].
\label{eq-decomp}
\end{equation}
Using Eqs.~(\ref{eq-inverse1}) and (\ref{kernel-mono}), we find after
some simple algebra  
\begin{equation}  
[1]=-\int_{-\infty}^0 d\nu\,
j(\nu,\xi)\frac{1}{2}\int_1^\infty\frac{d\eta}{\eta}\,\frac{\ue
  ^{-\eta\xi\tau}}{\eta\xi-\nu},
\label{eq-coherent1}
\end{equation}
\begin{equation}
[2]=\int_0^{\infty} d\nu\,
j(\nu,\xi)\frac{1}{2}\int_1^\infty\frac{d\eta}{\eta}\,\frac{\ue
  ^{-\eta\xi\tau}-\ue ^{-\nu\tau}}{\nu -\eta\xi},
\label{eq-coherent2}
\end{equation}
\begin{equation}  
 [3]=\int_0^{\infty} d\nu\,
j(\nu,\xi)\frac{1}{2}\int_1^\infty\frac{d\eta}{\eta}\,\frac{\ue
  ^{-\nu\tau}}{\nu +\eta\xi}.
\label{eq-coherent3}
\end{equation}
In [2], we separate the two terms. This requires taking the two new
integrals in principal part.  Regrouping 
the first one with [1] and the second one with [3], we obtain
\begin{eqnarray}  
& &  [1] +[2] +[3] = \int_0^{\infty} d\nu\, \ue
  ^{-\nu\tau}j(\nu,\xi)\frac{1}{2}\pvint_1^\infty\frac{d\eta}{\eta}\left[
\frac{1}{\eta\xi +\nu} + \frac{1}{\eta\xi -\nu}\right]\nonumber\\
 & & + \ \int_{-\infty}^{+\infty} d\nu\, 
j(\nu,\xi)\frac{1}{2}\pvint_1^\infty\frac{d\eta}{\eta}\frac{\ue
  ^{-\eta\xi\tau}}{\nu -\eta\xi}.
\label{eq-total}
\end{eqnarray}
The symbol $\pvint$ indicates that the integral is taken in principal part.
The first integral in the right-hand side has obviously the structure
of a Laplace transform. Making in the second integral the change of
variable $\nu=\nu'$ and $\eta\xi=\nu$ and interchanging the order of
the integration, we can write
\begin{equation}  
\int_{-\infty}^{+\infty} d\nu\, 
j(\nu,\xi)\frac{1}{2}\pvint_1^\infty\frac{d\eta}{\eta}\frac{\ue
  ^{-\eta\xi\tau}}{\nu -\eta\xi}= \int_\xi^\infty\frac{d\nu}{2\nu}\ue
  ^{-\nu\tau}\pvint_{-\infty}^{+\infty} d\nu'\,\frac{j(\nu',\xi)}{\nu'-\nu}. 
\label{eq-transform}
\end{equation}
The second integral has thus also the structure of a Laplace
transform. We thus have
\begin{eqnarray}  
& &  [1] +[2] +[3] = La\biggl\{j(\nu,\xi)\pvint_0^\infty k(\nu',\xi)\left[
\frac{1}{\nu' +\nu} + \frac{1}{\nu' -\nu}\right]\,d\nu'\nonumber\\
 & & + \ k(\nu,\xi)\pvint_{-\infty}^{+\infty}\frac{j(\nu',\xi)}{\nu'
    -\nu}\,d\nu'\biggr\}, 
\label{eq-final}
\end{eqnarray}  
where $La$ stands for Laplace transform and
\begin{equation}  
  k(\nu,\xi)\equiv 1/(2\nu),\ \  \nu>\xi,\quad k(\nu,\xi)\equiv 0,\ \ 0<\nu<\xi.
\label{eq-defknuapp}
\end{equation}
Proceeding in the same way with the second integral in the right-hand
side of Eq.~(\ref{eq-avintbis2}) and replacing in the left-hand side ${\cal
  J}(\tau,\xi)$  by $La\{j(\nu,\xi)\}$, we obtain for
$j(\nu,\xi)$ the singular integral equation in Eq.~(\ref{eq-cauchy2}).

There is a large choice of methods to transform a radiative
transfer equation or a convolution integral equation into a singular
integral equation. Instead of assuming that the
radiation field can be represented by a sum of exponentials, which the
hypothesis behind the inverse Laplace transform method
and the Case expansion method, one can
apply a direct Laplace transform to the radiative transfer equation or
to the associated convolution integral equation. For example a direct Laplace
transform applied to Eq.~(\ref{eq-avintbis2}) leads to  a singular
integral equation adjoint to Eq.~(\ref{eq-cauchy2}), in which the
coefficient $k(\nu,\xi)$ is inside the singular integral and the
inhomogeneous term is represented by its direct Laplace transform. The
required algebra is very similar to the algebra described above
\citep[e.g.][]{kour63,frisch22}. One of the first application of a
Laplace transform to construct an exact solution for a semi-infinite
medium can be found in \citet{halpern38}. The resulting singular
integral equation is solved by a factorization method \`a la Wiener--Hopf.
Direct and inverse Laplace transforms have also been employed to
developed accurate numerical methods such as the $F_N$ method or the
discrete-ordinate method \citep[e.g.][]{barichello98,barichello99b}.
The direct and inverse Laplace transform methods lead to singular
integral equations for infinite as well as semi-infinite media. For an
infinite medium, a Fourier transform can provide an algebraic
equation. For a semi-infinite medium, it can  also lead
to a singular integral equation. A very elegant proof, making use of the
analyticity properties of the complex Fourier transform of the
Wiener--Hopf integral operator, is described in \citet{pincus66}. We
recall that the traditional Wiener--Hopf method leads to a
factorization problem in the complex plane.

\section{The generalized dispersion function $\Omega(z)$}
\label{sec-omega}

The generalized dispersion function $\Omega(z)$ is defined in the text by
\begin{equation}  
  \Omega(z)\equiv 1-\omega(1-b)\int_0^{\varphi_0}W(\xi)\xi
  \frac{T(z,\xi)}{{\cal L}(z,\xi)}\,d\xi.
  \label{def-omega4}
\end{equation}
This definition is used in Section~\ref{sec-omega1} to
analyze the properties of $\Omega(z)$ and in Section~\ref{sec-omega2}
to show that $\Omega(z)$ reduces to the complete frequency
redistribution dispersion function ${\cal L}_{\rm R}(z)$ when
$b=0$. We recall that $\xi=\varphi(x)$, with $\varphi(x)$ the
absorption profile and $\varphi_0=\varphi(0)$.
\begin{figure}
\begin{center}
\includegraphics[scale=0.6]{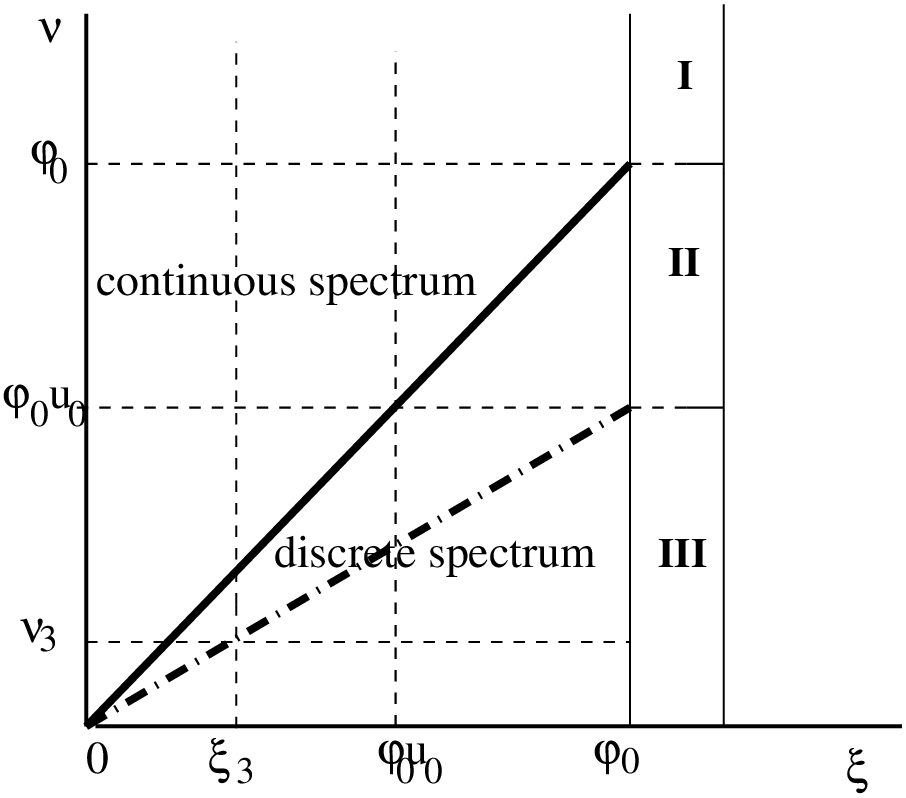}
\caption{Properties of the spectrum of ${\cal L}(\nu,\xi)$ in the
  variables $\nu$ 
and $\xi$. The line $\nu=\xi$ shows the separation between
the continuous and the discrete spectrum. The dotted-dashed line
$\nu=u_0\xi$ shows 
the location of the zeroes of ${\cal L}(z,\xi)$, with $u_0$ the
positive zero of $\tilde{\cal L}(\tilde z)$. The domains {\bf
  I}, {\bf II}, and {\bf III} corresponds respectively to
$\nu>\varphi_0$, $\varphi_0u_0<\nu<\varphi_0$ and
$0<\nu<\varphi_0u_0$. $\nu_3$ is a typical values of
$\nu$ in the domain {\bf III} and $\xi_3$ is the value of $\xi$ such
that the positive zero of ${\cal L}(z,\xi)$ is located at
$\nu_3$. Integrations are along lines parallel to the horizontal
$\xi$-axis for the calculation of $\Omega ^\pm(\nu)$ and 
along lines parallel to the vertical $\nu$-axis for the calculations of
${\cal J}(\tau,\xi)$ and $\bar{\cal J}(\tau)$.}
\label{figure1}
\end{center}
\end{figure}
\subsection{Properties of $\Omega(z)$}
\label{sec-omega1} 

We first remark that $\Omega(z)$ is analytic in the complex plane cut
along the real axis, the point zero being excluded. Indeed, it has the same
singular line as ${\cal L}(z,\xi)$,  and in the integration over
$\xi$, the lower limit of $\xi$ is zero. For $z=0$, we have ${\cal
  L}(0,\xi)=\epsilon_{b}$ (see Eq.~(\ref{eq-lzero2})). Using
${\cal L}(0,\xi)=1-\omega\,b\xi T(0,\xi)$ and  the normalization of
$W(\xi)$ to unity (see Eq.~(\ref{eq-normewxi})), we readily infer from
Eq.~(\ref{def-omega4}) that
\begin{equation}  
\Omega(0)=\frac{\epsilon}{\epsilon_{b}}.
\label{eq-omegazero}
\end{equation}
Hence $\Omega(0)=\epsilon$ for $b=0$. For $b=1$, $\Omega(z)$ is equal
to 1 for all $z$.  At infinity, ${\cal L}(z,\xi)\to 1$ and
$T(z,\xi)\to0$, hence $\Omega(z)$ tends to 1.

We now determine for all $\nu\in]-\infty,+\infty[$ ($\nu=\Re(z)$) the
  limiting values $\Omega^+(\nu)$ and $\Omega^-(\nu)$ of $\Omega(z)$,
  as $z$ tends to the real axis, from above and from below. With the
  explicit expressions of $\Omega^\pm(\nu)$, it will be possible to
  determine whether or not $\Omega(z)$ is free of zeroes. We know that
  the dispersion function ${\cal L}(\nu,\xi)$ has a continuous spectrum
  for $|\nu|>|\xi|$ and a discrete one for $\nu\in[-\xi,+\xi]$,
  with zeroes at $\pm\nu_0(\xi)$. The existence of these two different
  spectra and of these zeroes brings some complexity to the
  calculation of $\Omega^\pm(\nu)$ and the expressions that are
  obtained depend on the value of $\nu$.

  Three different domains shown in Fig.~\ref{figure1} have to be
  considered. They are referred to 
  as {\bf I}, for $\nu$ larger that $\varphi_0$, {\bf II}, for
  $\varphi_0u_0<\nu<\varphi_0$, and {\bf III}, for
  $0<\nu<\varphi_0u_0$.  Because $\Omega(z)$ is an even function of
  $z$, we consider only positive values of $\nu$. The straight line
  $\nu=\xi$ separates the continuous from the  discrete spectrum. The
  line $\nu=u_0\xi$ shows the location of the zeroes of ${\cal
    L}(\nu,\xi)$. We recall that $u_0$ is the positive zero of
  $\tilde{\cal L}(\tilde z)$ (see Eq.~(\ref{eq-dispersion3})) and that
$u_0$ and $\varphi_0$ are smaller than one.\\

\noindent{\bf Domain I} $\nu\ge\varphi_0$\\

For all values of $\nu$ in this interval and all values of
$\xi\in[0,\varphi_0]$, 
$\nu$ belongs to ${\cal C}(\xi)$, that is to the continuous
spectrum. The Plemelj formulae can be applied to
Eq.~(\ref{def-omega4}).  Using the expressions of ${\cal L}^\pm(\nu)$
and $T^\pm(\nu)$ given in Eq.(\ref{dispplus}) to
calculate the jump of $T(z,\xi)/{\cal L}(z,\xi)$, we obtain
\begin{equation}  
\Omega ^+_{\rm I}(\nu)- \Omega
^-_{\rm I}(\nu)=-\omega(1-b)2\ui\pi\int_0^{\varphi_0}W(\xi)\xi
\frac{k(\nu,\xi)}{D(\nu,\xi)}\,d\xi,
\label{eq-omegajump2}
\end{equation}  
and 
\begin{equation}  
\Omega^+_{\rm I}(\nu)+\Omega^-_{\rm I}(\nu)=2-\omega(1-b)2
\int_0^{\varphi_0}W(\xi)\xi \frac{C(\nu,\xi)}{D(\nu,\xi)} \,d\xi,
\label{eq-omegasomme2}
\end{equation}
where
\begin{equation}  
C(\nu,\xi)=t(\nu,\xi) - \omega\,b\xi E(\nu,\xi),
\label{def-grandC}
\end{equation}
\begin{equation}  
E(\nu,\xi)=T^+(\nu,\xi)T^-(\nu,\xi)=t^2(\nu,\xi)+ \pi^2k^2(\nu,\xi),
\label{def-grandE}
\end{equation}
and 
\begin{equation}  
D(\nu,\xi)={\cal L}^+(\nu,\xi){\cal L}^-(\nu,\xi)=\lambda
^2(\nu,\xi) +\pi ^2\omega^2b^2\xi ^2\,k^2(\nu,\xi).
\label{def-productpm}
\end{equation}
We thus obtain
\begin{eqnarray}
& &   \Re[\Omega^\pm_{\rm I}(\nu)]=1-\omega(1-b)\int_0^{\varphi_0}W(\xi)\xi
  \frac{C(\nu,\xi)}{D(\nu,\xi)}\,d\xi,\nonumber\\
& &  \Im[\Omega^\pm_{\rm I}(\nu)]=\mp\pi \omega(1-b)\int_0^{\varphi_0}W(\xi)\,\xi
\frac{ k(\nu,\xi)}{D(\nu,\xi)}\,d\xi.
\label{eq-omegapm1}
\end{eqnarray}

\noindent{\bf Domain II} $u_0\varphi_0<\nu\le\varphi_0$\\

For all $\xi<\nu$, $\nu$ belongs to the continuum spectrum and for all
$\xi>\nu$, to the discrete spectrum. In the latter region, ${\cal
  L}(z,\xi)$ being analytic, its jump across the real axis is zero. In
domain {\bf II} there is no contribution from the zero at $\nu_0(\xi)$
 since $\nu>u_0\varphi_0$ (see Fig.~\ref{figure1}). We thus obtain 
\begin{eqnarray}  
& &   \Re[\Omega^\pm_{\rm II}(\nu)]=1-\omega(1-b)\left[\int_0^{\nu}W(\xi)\xi
\frac{C(\nu,\xi)}{D(\nu,\xi)}\,d\xi
+ \int_\nu^{\varphi_0}W(\xi)\xi\frac{t(\nu,\xi)}
{\lambda(\nu,\xi)}\,d\xi\right],\nonumber\\   
& & \Im[\Omega^\pm_{\rm II}(\nu)]=\mp\pi\omega(1-b)\int_0^{\nu}W(\xi)\xi
\frac{k(\nu,\xi)}{D(\nu,\xi)}\,d\xi.  
\label{eq-omegapm2}
\end{eqnarray}
In the imaginary part, which comes from the continuous spectrum, the
upper limit of the integral is now $\nu$.\\  

\noindent{\bf Domain III} $0<\nu<\varphi_0u_0$\\

The situation similar to that of the domain {\bf II}, but for a given
value of $\nu=\Re(z)$, ${\cal L}(z,\xi)$ will have a zero when $\xi$
has the value $\xi_0(\nu)=\nu/u_0$ (see
Section~\ref{sec-dispersionfreq}). There is thus in
Eq.~(\ref{eq-omegapm2}) an additional contribution coming from the
integral
\begin{equation}  
\int_\nu^{\varphi_0}W(\xi)\xi \frac{T(\nu,\xi)}{{\cal
    L}(\nu,\xi)}\,d\xi. 
\label{eq-integrale}
\end{equation}
The analyticity of
$T(z,\xi)$ and ${\cal L}(z,\xi)$ for $\Re(z)\in[-\xi,+\xi]$, allows us
to deform the integration line in such a way that it will pass around
the pole. The contribution of the pole is found, in the standard way,
by introducing a semi-circle around it,
placed below the integration line for $\Omega ^+(\nu)$ and above it for
$\Omega^-(\nu)$. Taylor expanding $\lambda(\nu,\xi)$ around $\xi_0(\nu)$, we
find that the contribution of the pole to the integral in
Eq.~(\ref{eq-integrale}) may be written as
\begin{equation}  
 \pm \ui\pi
 W(\xi_0(\nu))\xi_0(\nu)\frac{t(\nu,\xi_0(\nu))}{\partial_\xi
   {\lambda}(\nu,\xi_0(\nu))}, 
\label{eq-temp}
\end{equation} 
where $\partial_\xi {\lambda}(\nu,\xi_0)$ is the partial derivative of
${\lambda}(\nu,\xi)$ with respect to $\xi$ taken at $\xi_0(\nu)$. This
partial derivative has been calculated in
Section~\ref{sec-dispersionfreq} (see Eq.~(\ref{eq-derivativexi})).
We thus find
\begin{equation}  
\Omega^\pm_{\rm III}(\nu)= \Omega^\pm_{\rm II}(\nu)
  \mp\ui\pi\omega(1-b)\,W(\xi_0(\nu))\xi_0(\nu)
  \frac{t(\nu,\xi_0(\nu))}{\partial_\xi {\lambda}(\nu,\xi_0(\nu))}.
\label{eq-omegapm3}
\end{equation}
A general theory for  the calculation of integrals with a pole in the
denominator as in Eq.~(\ref{eq-integrale})  can be
found in \citet[p.~66-68]{gakhov66} and in \citet{hemsch72}.

\begin{figure}
\begin{center}
\includegraphics[scale=0.35]{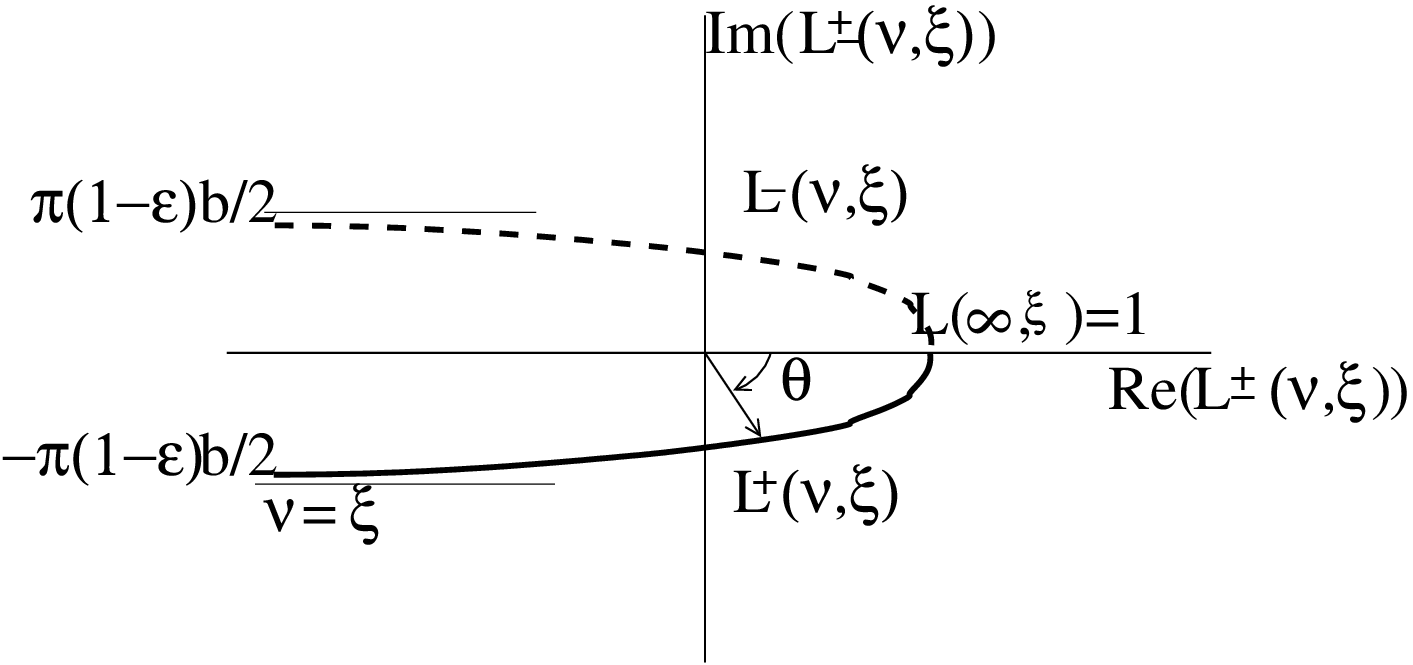}
\includegraphics[scale=0.35]{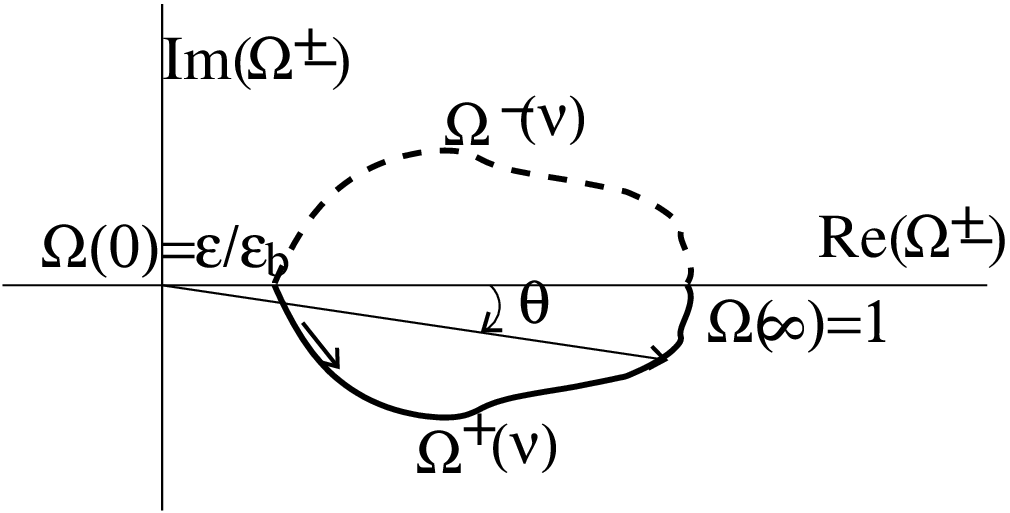}
\end{center}
\vspace{-0.2cm}
\caption{Phase diagrams of the frequency dependent dispersion
  function ${\cal L}(z,\xi)$ and of the generalized dispersion
  function $\Omega(z)$. The upper
panel shows $\Im[{\cal L}^\pm(\nu,\xi)]$ versus $\Re[{\cal L}^\pm(\nu,\xi)]$
for $\nu\in[\xi,\infty[$  and the lower panel $\Im[\Omega ^\pm(\nu)]$
    versus $\Re[\Omega ^\pm(\nu)]$ for $\nu\in[0,\infty[$.}
\label{figure2}
\end{figure}

We now turn to the determination of the number of zeroes of
$\Omega(z)$. A standard method for an analytic function is to examine
the variation of its argument along a close contour inside the
analyticity region \citep[e.g.,][]{carrier66,casezwei67}.
For $\Omega(z)$, which is an even function of $z$, it suffices to
calculate the variation of the argument of $\Omega^+(\nu)$ as $\nu$
varies along the positive real axis. A phase diagram of $\Omega
^+(\nu)$, showing the variation of $\Im[\Omega ^+(\nu)]$ versus
$\Re[\Omega ^+(\nu)]$ for $\nu\in[0,\infty[$ is sketched in
    Figure~\ref{figure2}. For $\nu=0$, we know that
    $\Omega(0)=\epsilon/\epsilon_{b}$ (see Eq.~(\ref{eq-omegazero}))
    hence $\Im[\Omega ^+(\nu)]=0$ and $\Re[\Omega ^+(\nu)]$ is smaller
    or equal to 1. At infinity $\Omega(z)\to 1$, hence, $\Im[\Omega
      ^+(\infty)]=0$ and $\Re[\Omega ^+(\infty)]=1$.
    Equations~(\ref{eq-omegapm1}) and (\ref{eq-omegapm2}) show that
    $\Im[\Omega ^+(\nu)]$ is negative in domains {\bf I} and {\bf
      II}. In domain {\bf III}, the imaginary part in
    Eq.~(\ref{eq-omegapm3}) contains an additional term coming from
    the pole. Since the derivative of ${\cal L}(\nu,\xi)$ is positive,
    as pointed in Section~\ref{sec-dispersionfreq}, this additional
    term is also negative. The real part $\Re[\Omega ^+(\nu])$ is
    positive for $\nu=0$ and at infinity. It can be checked that it
    remains positive for all $\nu$ positive. From these properties,
    illustrated in the lower panel of Fig.~\ref{figure2}, one infers
    that $(\theta(\infty)-\theta(0))$, the variation of the argument
    of $\Omega ^+(\nu)$ is zero, hence that $\Omega(z)$ is free of
    zeroes. The upper panel shows for comparison a sketch of the phase
    diagram of ${\cal L}(\nu,\xi)$.

\subsection{The complete frequency redistribution limit}
\label{sec-omega2}
We verify in this appendix that
$\Omega(z)={\cal L}_{\rm R}(z)$ for  $b=0$, with ${\cal L}_{\rm R}(z)$
the dispersion function for complete frequency redistribution. 
In this limit the source function depends only on $\tau$ and satisfies
a convolution integral equation, given in Eq.~(\ref{eq-convolrc}) for
$\tau\in[0,\infty[$. The kernel, defined by
\begin{equation}  
K(\tau)= \frac{1}{2}\int_{-\infty}^{\infty}
\int_0^1\varphi^2(x)\ue^{-|\tau|\varphi(x)/\mu}
\frac{d\mu}{\mu}\,dx,
\label{eq-appkernelrc}
\end{equation}
may be written as
\begin{equation}  
  K(\tau)=\int_0^\infty k_{\rm R}(\nu)\,\ue^{-\nu|\tau|}\,d\nu.
\label{eq-kernelc2}
\end{equation}
Its inverse Laplace transform $k_{\rm R}(\nu)$ (R for complete
frequency redistribution) may be written as
\begin{equation}  
  k_{\rm R}(\nu)=\frac{1}{\nu}\int_{y(\nu)}^\infty\varphi ^2(x)\,dx,
\label{eq-knurc}
\end{equation}
where
\begin{equation}  
  y(\nu)=\left\{\begin{array}{cc}
\varphi^{-1}(\nu) & \nu<\varphi_0\\
0 & \nu>\varphi_0.\end{array}\right.
\label{limitinf}
\end{equation}
The notation $\varphi^{-1}$ stands for the inverse function of
$\varphi$. The dispersion function is then given by  
\begin{equation}  
  {\cal L}_{\rm R}(z)\equiv 1-\omega\int_0^\infty k_{\rm R}(\nu)
\left[\frac{1}{\nu-z} +\frac{1}{\nu+z}\right] \,d\nu.
\label{eq-disprc}
\end{equation}
The integral in the right-side is the complex Fourier transform of the
kernel.

The frequency dependent dispersion function ${\cal L}(z,\xi)$,
reduces to ${\cal L}(z,\xi)=1$ when $b=0$ (see
Eq.~(\ref{eq-dispersion})). Hence $\Omega(z)$ reduces to 
\begin{equation}  
\Omega(z)=1-\omega
\int_0^{\varphi_0}W(\xi)\,\xi\,T(z,\xi)\,d\xi,  
\label{eq-omegacr}
\end{equation}
with
\begin{equation}  
 T(z,\xi)=\int_{\xi}^\infty k(\nu,\xi)\left[\frac{1}{\nu -z} + \frac{1}{\nu
     +z}\right]\,d\nu, 
\label{eq-defT2}
\end{equation}
and
\begin{equation}  
  k(\nu,\xi)=\frac{1}{2\nu}Y(\nu-\xi).
\label{}
\end{equation}
The function $Y(\nu)$ is the Heaviside function.  Replacing $\xi$ by
$\varphi(x)$ and using
\begin{equation}  
 W(\xi)=2\xi|\frac{dx}{d\xi}|,  
\label{eq-weight}
\end{equation}
we can rewrite Eq.~(\ref{eq-omegacr}) as
\begin{equation}  
 \Omega(z)= 1-\omega\int_0^{\infty}\varphi ^2(x)\int_{\varphi(\nu)}^\infty
\frac{1}{\nu}Y(\nu-\varphi(x))\frac{2\nu}{\nu ^2 -z^2}\,d\nu\,dx.    
\label{eq-omegacr2}
\end{equation}
Interchanging the order of the integrations between $\nu$ and $x$
and using $Y(\nu-\varphi(x))=1$, we obtain
\begin{equation}  
\Omega(z)=1-\omega \int_0^{\infty}d\nu\left[\frac{1}{\nu -z} +
  \frac{1}{\nu +z}\right]\frac{1}{\nu}\int_{y(\nu)}^\infty \varphi
^2(\lambda)\,d\lambda.   
\label{eq-omegacr3}
\end{equation}
Referring to the definition of $k_{\rm R}(\nu)$ given
in Eq.~(\ref{eq-knurc}), we can   
immediately verify that $\Omega(z)={\cal L}_{\rm R}(z)$.

A similar calculation, applied to the expressions given in
Section~\ref{sec-omega1}  for $\Omega ^+(\nu)-\Omega ^-(\nu)$ show 
that in the limit  $b=0$,
\begin{equation}  
\Omega ^+(\nu)-\Omega ^-(\nu)=-\omega 2\ui\pi\,
  k_{\rm R}(\nu), \quad \nu=\Re(z)\in]-\infty,+\infty[.
\label{eq-jumplcr}
\end{equation}
This is precisely the jump of ${\cal L}_{\rm R}(z)$, as can be derived from
Eq.~(\ref{eq-omegacr3}). 

\begin{figure}
\begin{center}
\includegraphics[scale=0.45]{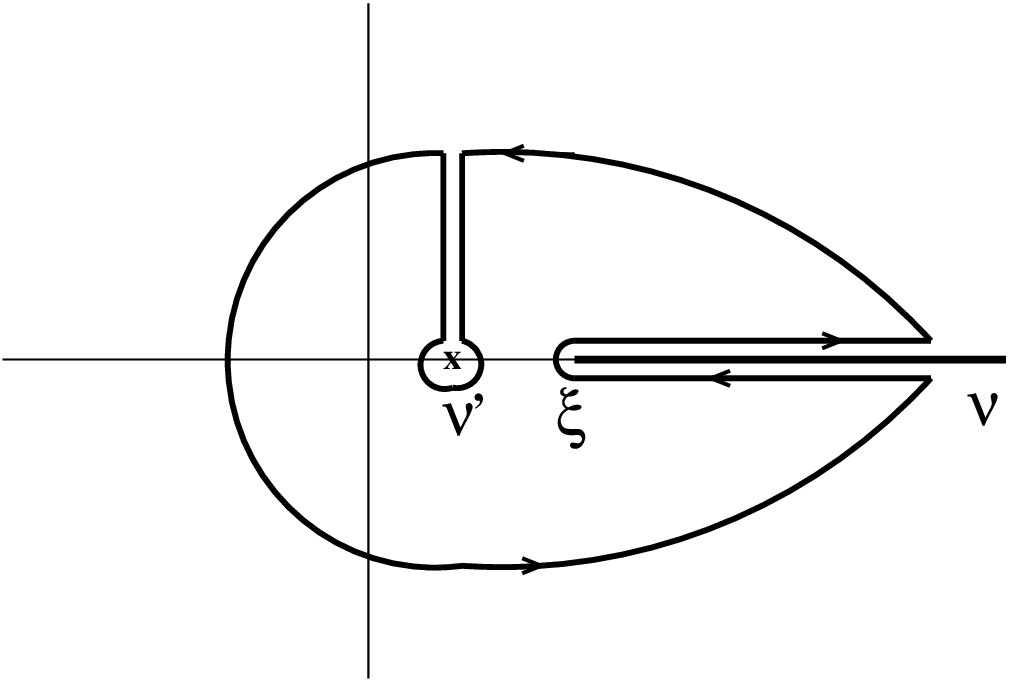}
\end{center}
\caption{Contour in the complex plane for the calculation of the
  integral in Eq.~(\ref{eq-int1}). The function $X(z,\xi)$ is analytic in the
  complex plane cut along the semi-infinite interval
  $[\xi,\infty[$. The singularity of the integrand is at $\nu'$, with
      $\nu'\in[0,\xi]$.} 
\label{fig-s2-contour}
\end{figure}
\section{The functions  ${\cal M}(z,\xi)$ and ${\cal V}(\nu,\nu')$}
\label{sec-jump}

The function ${\cal M}(z,\xi)$ is defined by
\begin{equation}  
  {\cal M}(z,\xi)\equiv H_{\xi}^\infty\left[
\frac{[X^\ast]^+(\nu,\xi)}{{\cal L}^+(\nu,\xi)}\,m(\nu,\xi)\right],
\label{eq-defS1S2app}
\end{equation}
where $H_{\xi}^\infty$ stands for the Hilbert transform over
$[\xi,\infty[$ and
\begin{equation}  
 m(\nu,\xi)= p^\ast(\nu)
- \left[{\cal L}^+(\nu,\xi) - {\cal
      L}^-(\nu,\xi)\right] \frac{1}{2\ui\pi}\pvint_{0}^{\xi}
\frac{p^\ast(\nu')}{\lambda(\nu',\xi)}\frac{d\nu'}{\nu'-\nu}.
\label{eq-semipmnuxi2}
\end{equation}
For convenience we write
${\cal M}(z,\xi)= {\cal M}_1(z,\xi) + {\cal M}_2(z,\xi)$  with 
\begin{equation}
{\cal M}_2(z,\xi)=-\frac{1}{2\ui\pi}\int_{\xi}^\infty\biggl\{
  \left[[X^\ast]^+(\nu,\xi) -[X^\ast]^-(\nu,\xi)\right]
\frac{1}{2\ui\pi}\pvint_{0}^{\xi}
  \frac{p^\ast(\nu')}{\lambda(\nu',\xi)}\frac{d\nu'}{\nu'-\nu}\biggr\}\, 
\frac{d\nu}{\nu-z}. 
\label{eq-grands2}
\end{equation}
This double integral can be transformed into the sum of two simple
integrals. Using 
\begin{equation}  
  \frac{1}{\nu'-\nu}\frac{1}{\nu -z}=\left[\frac{1}{\nu'-\nu} +
    \frac{1}{\nu -z}\right]\frac{1}{\nu'-z}, 
\label{}
\end{equation}
and changing the order of integration, we find
\begin{eqnarray}  
& & {\cal M}_2(z,\xi)=-\frac{1}{2\ui\pi}\pvint_{0}^{\xi}
  \frac{p^\ast(\nu')}{\lambda(\nu',\xi)}\frac{d\nu'}{\nu'-z}\nonumber\\ 
& & \times \frac{1}{2\ui\pi}\int_{\xi}^\infty 
  \Bigl\{[X^\ast]^+(\nu,\xi)-[X^\ast]^-(\nu,\xi)\Bigr\}
  \left[\frac{1}{\nu'-\nu} + \frac{1}{\nu -z}\right]\,d\nu,
\label{eq-grands2bis}
\end{eqnarray}
where $X^\ast(z,\xi)=(\nu_0(\xi)-z)X(z,\xi)$. The integration
contour shown in Fig.~\ref{fig-s2-contour} leads to
\begin{equation}  
\frac{1}{2\ui\pi}\int_{\xi}^\infty \Bigl\{[X^\ast]^+(\nu,\xi)
  -[X^\ast]^-(\nu,\xi)\Bigr\}\,\frac{d\nu}{\nu-\nu'}= X^\ast(\nu',\xi) -1.  
\label{eq-int1}
\end{equation}
A similar equation holds when $\nu'$ is replaced by a point $z$ in the
complex plane. Summing the two equations, we obtain
\begin{equation}
{\cal M}_2(z,\xi)=-X^\ast(z,\xi)\frac{1}{2\ui\pi}\pvint_{0}^{\xi}
\frac{p^\ast(\nu')}{\lambda(\nu',\xi)}\,\frac{d\nu'}{\nu'-z}
+ \frac{1}{2\ui\pi}\pvint_{0}^{\xi}\frac{p^\ast(\nu')}
       {\lambda(\nu',\xi)}X^\ast(\nu',\xi)\,\frac{d\nu'}{\nu'-z}.
\label{eq-grands2f}
\end{equation}
We now use the factorization
$X^\ast(\nu,\xi)X^\ast(-\nu,\xi)=\lambda(\nu,\xi)$, which holds for
$\nu\in[0,\xi]$, to transform the second integral in the right-hand
side of Eq.~(\ref{eq-grands2f}). To calculate ${\cal M}_1(z,\xi)$, we
use the factorization $[X^\ast]^+(\nu,\xi)X^\ast(-\nu,\xi)={\cal
  L}^+(\nu,\xi)$, which holds for $\nu\in[\xi,\infty[$. We thus
finally obtain
\begin{equation}  
{\cal M}(z,\xi)=-\frac{1}{2\ui\pi}X^\ast(z,\xi)\pvint_{0}^{\xi}
  \frac{p^\ast(\nu')}{\lambda(\nu',\xi)}\,\frac{d\nu'}{\nu'-z}
+ \frac{1}{2\ui\pi}\pvint_{0}^{\infty}
  \frac{p^\ast(\nu')}{X^\ast(-\nu',\xi)}\,\frac{d\nu'}{\nu'-z}.
\label{eq-grandmcurl}
\end{equation}

In Section~\ref{sec-fredholm} we show that $\rho(\nu)$, the jump of
$R(z)$, can be written as
\begin{equation}  
  \rho(\nu)=\frac{1-b}{b}\pvint_0^\infty {\bar\jmath}(\nu'){\cal
    V}(\nu,\nu')\,d\nu', 
\label{eq-finalrhoapp}
\end{equation}
where ${\cal V}(\nu,\nu')$  is the
jump across the positive real axis of the function 
\begin{equation}  
  V(z,\nu)\equiv\frac{1}{2\ui\pi}\int_0^{\varphi_0}\frac{W(\xi)}{{\cal
    L}(z,\xi)}r(z,\nu,\xi)\,d\xi.
\label{eq-defvznu2}
\end{equation}
To calculate ${\cal V}(\nu,\nu')$ we can
proceed exactly as for the calculation the jump of $\Omega(z)$ in
Appendix~\ref{sec-omega1}. In both cases, the jumps are 
controlled by the jump of $1/{\cal L}(z,\nu)$. We thus obtain
\begin{equation}  
 {\cal V}(\nu,\nu')= V(\nu,\nu'),\quad \nu\in[u_0\varphi_0,\infty[,
\label{eq-defcalv1}
\end{equation}
\begin{equation}  
{\cal V}(\nu,\nu')= V(\nu,\nu')
+  r(\nu,\nu',\xi_0)\frac{W(\xi_0)}{\partial_\xi{\cal 
    L}(\nu,\xi_0)},\, \nu\in[0,u_0\varphi_0], 
\label{eq-defcalv2}
\end{equation}
where
\begin{equation}  
  V(\nu,\nu')\equiv\frac{1}{2\ui\pi}\int_0^{\varphi_0}W(\xi)\left[\frac{1}
{{\cal L}(z,\xi)}\right]_{\rm jmp}r(\nu,\nu',\xi)\,d\xi, 
\label{eq-defgrv}
\end{equation}
and
\begin{equation}  
  r(\nu,\nu',\xi_0)=\left[\frac{2\nu
X(-\nu,\xi_0)} {(\nu + \nu')X(-\nu',\xi_0)}-1\right]\frac{P}{\nu'-\nu}.
\label{eq-defpetitrzero}
\end{equation}
Here, $\xi_0=\nu/u_0$, where $u_0$ is the positive zero of the
function $\tilde{\cal L}(\tilde z)$ introduced in
Eq.~(\ref{eq-dispersion3}). In the complete frequency redistribution
limit ${\cal V}(\nu,\nu')$ and $\rho(\nu)$ are zero.

\end{appendix} 
\end{document}